\documentclass[twocolumn,epjc3]{svjour3mod}  
\RequirePackage[T1]{fontenc}
\RequirePackage{psfrag}
\RequirePackage{graphicx}
\RequirePackage{mathptmx}      
\RequirePackage{flushend}
\RequirePackage{latexsym}
\RequirePackage[colorlinks,citecolor=blue,urlcolor=blue,linkcolor=blue]{hyperref}
\RequirePackage{amsmath,amssymb}
\RequirePackage[normalem]{ulem}
\RequirePackage{booktabs}
\RequirePackage{xcolor}
\RequirePackage{braket}
\usepackage[utf8]{inputenc}
\usepackage{siunitx}
\usepackage{cite}
\definecolor{tcol}{RGB}{45,181,86}

\title{Towards an improved understanding of {$\boldsymbol{\eta\to\gamma^*\gamma^*}$}}

\author{S.~Holz\thanksref{addrBN1,addrBN2,e6} \and
J.~Plenter\thanksref{addrBN1,addrBN2,addrVal}
\and
C.~W.~Xiao\thanksref{addrJ1,addrChangsha}
\and
T.~Dato\thanksref{addrBN1} 
\and
C.~Hanhart\thanksref{addrJ1}
\and
B.~Kubis\thanksref{addrBN1,addrBN2,e2} 
\and
U.-G.~Mei{\ss}ner\thanksref{addrBN1,addrBN2,addrJ1,addrGeorg} 
\and
A.~Wirzba\thanksref{addrJ1}
}

\thankstext{e6}{e-mail: holz@hiskp.uni-bonn.de}
\thankstext{e2}{e-mail: kubis@hiskp.uni-bonn.de}

\institute{Helmholtz-Institut f\"ur Strahlen- und Kernphysik (Theorie), 
           Universit\"at Bonn, 53115  Bonn, Germany \label{addrBN1}
           \and
           Bethe Center for Theoretical Physics,
           Universit\"at Bonn, 53115  Bonn, Germany \label{addrBN2}
           \and
           Instituto de F\'isica Corpuscular, Universitat de Val\`encia -- Consejo Superior de Investigaciones
           Científicas, Parc Científic, 46980 Paterna, Valencia, Spain \label{addrVal}
           \and
           Institut  f\"{u}r Kernphysik (Theorie), 
           Institute for Advanced Simulation, and
           J\"ulich Center for Hadron Physics,
           Forschungszentrum J\"ulich,  \\
           52425 J\"{u}lich, Germany  \label{addrJ1}
           \and
           School of Physics and Electronics, Central South University, Changsha 410083, China \label{addrChangsha}
           \and
           Tbilisi State  University,  0186 Tbilisi, Georgia \label{addrGeorg}
}

\newcommand{\diff}{\text{d}}
\newcommand{\F}{\mathcal{F}}
\newcommand{\eps}{\epsilon}
\newcommand{\GeV}{\,\text{GeV}}
\newcommand{\MeV}{\,\text{MeV}}

\newcommand{\bsp}{\begin{sloppypar}}
\newcommand{\esp}{\end{sloppypar}}

\begin{document}

\date{}
\maketitle

\begin{abstract}\bsp
We argue that high-quality data on the reaction $e^+e^-\to \pi^+\pi^-\eta$ will allow one to determine
the doubly-virtual form factor $\eta\to \gamma^*\gamma^*$ in a model-independent way
with controlled accuracy. This is an important step towards a reliable evaluation
of the hadronic light-by-light scattering contribution to the anomalous
magnetic moment of the muon. When analyzing the existing data for $e^+e^-\to \pi^+\pi^-\eta$ for total energies squared $k^2>1\GeV^2$,
we demonstrate that the effect of the $a_2$ meson provides a natural breaking mechanism for the commonly employed factorization ansatz in the doubly-virtual form factor $F_{\eta\gamma^*\gamma^*}(q^2,k^2)$. 
However, better data are needed to draw firm conclusions.
\esp
\end{abstract}

\section{Introduction}
\label{intro}

\begin{figure} 
 \centering
   \includegraphics[width=\columnwidth]{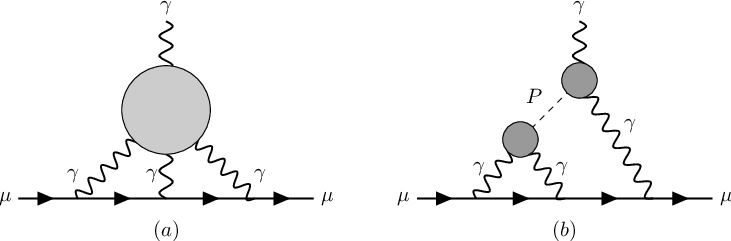}
   \caption{Diagrammatic representations of  the HLbL contribution to $(g-2)_\mu$ $(a)$, as well as pseudoscalar pole term contributions thereof $(b)$, where $P \in \lbrace \pi^0,\, \eta,\, \eta' \rbrace$. 
   } 
   \label{fig:g-2_hadLbL}
\end{figure}

\bsp
Transition form factors contain important information about the properties of 
the decaying particles. Additional interest in meson decays with one or two virtual
photons in the final state comes from the fact that the theoretical precision
of the Standard Model calculations for the anomalous magnetic moment of the muon is significantly affected by the one of hadronic light-by-light (HLbL) scattering, where the latter appears as
a sub-amplitude; see Fig.~\ref{fig:g-2_hadLbL}.
A few years ago, the Bern group reported on important progress towards a 
model-independent determination of the HLbL contribution based on dispersion theory~\cite{Colangelo:2014dfa,Colangelo:2014pva,Colangelo:2015ama}. 
In principle this allows for an analysis of similar rigor as commonly applied
for the hadronic vacuum polarization~\cite{Aoyama:2020ynm}.\footnote{See also the alternative approach to analyze the muon's Pauli form factor dispersively~\cite{Pauk:2014rfa}.}
The updated experimental value for $(g-2)_\mu$, combining the BNL measurement~\cite{Bennett:2006fi} with the first Fermilab results~\cite{Abi:2021gix,Albahri:2021ixb}, shows a $4.2\sigma$ tension as compared to the current theory consensus on the Standard Model value~\cite{Aoyama:2020ynm,Aoyama:2012wk,Aoyama:2019ryr,Czarnecki:2002nt,Gnendiger:2013pva,Davier:2017zfy,Keshavarzi:2018mgv,Colangelo:2018mtw,Hoferichter:2019gzf,Davier:2019can,Keshavarzi:2019abf,Hoid:2020xjs,Kurz:2014wya,Melnikov:2003xd,Masjuan:2017tvw,Colangelo:2017qdm,Colangelo:2017fiz,Danilkin:2021icn,Hoferichter:2018dmo,Hoferichter:2018kwz,Gerardin:2019vio,Bijnens:2019ghy,Colangelo:2019lpu,Colangelo:2019uex,Blum:2019ugy,Colangelo:2014qya}, and the upcoming improvement in the experimental statistics~\cite{Grange:2015fou} demands further efforts to reduce the theoretical uncertainty. 
Reference~\cite{Aoyama:2020ynm} identifies the $\eta$ and $\eta'$ pole terms as major outstanding contributions 
in the dispersive approach to HLbL, next to axial vectors~\cite{Hoferichter:2020lap,Zanke:2021wiq} and the matching to short-distance constraints~\cite{Melnikov:2003xd,Bijnens:2019ghy,Colangelo:2019lpu,Colangelo:2019uex,Knecht:2020xyr,Ludtke:2020moa,Bijnens:2020xnl,Bijnens:2021jqo}.  A refined understanding of the doubly-virtual $\eta$ transition form factor is the object of this article; see also the review Ref.~\cite{Gan:2020aco}.

In Ref.~\cite{Hanhart:2013vba} the isovector contribution of the
singly-virtual form factor $F_{\eta\gamma^*\gamma}(q^2) \equiv F_{\eta\gamma^*\gamma^*}(q^2,0)$
was calculated for virtualities $q^2\ll 1\GeV^2$ 
from data on $\eta\to\pi^+\pi^-\gamma$ and the pion vector form factor via a dispersion integral. 
This was done with the help of a convenient parameterization
of the corresponding $\pi\pi$ invariant-mass distribution derived in Ref.~\cite{Stollenwerk:2011zz}, 
see also Ref.~\cite{Adlarson:2011xb}. In particular,
it was demonstrated that when the high-statistics data of 
Ref.~\cite{Babusci:2012ft} were used to fix the $\eta\to\pi^+\pi^-\gamma$ input, this procedure leads to a determination of
$F_{\eta\gamma^*\gamma}(q^2)$ with an accuracy higher than that of 
the most recent direct measurements~\cite{Arnaldi:2009aa,Aguar-Bartolome:2013vpw,Arnaldi:2016pzu,Adlarson:2016hpp}.
Especially this is the case since the isoscalar contribution is negligibly small.

\begin{figure} 
 \centering
   \includegraphics[height=3.3cm]{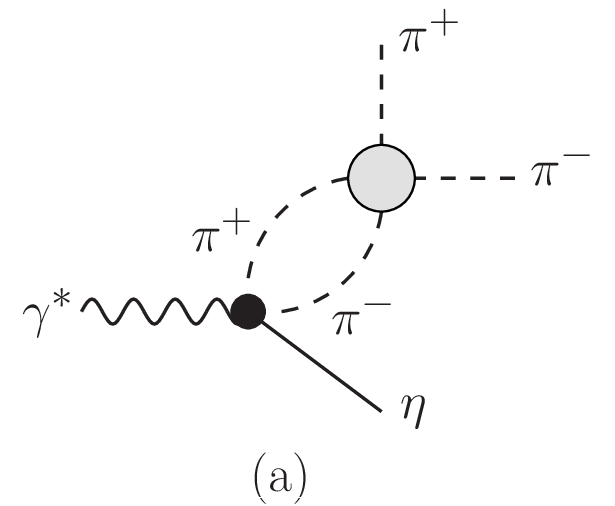} \hfill
   \includegraphics[height=3.3cm]{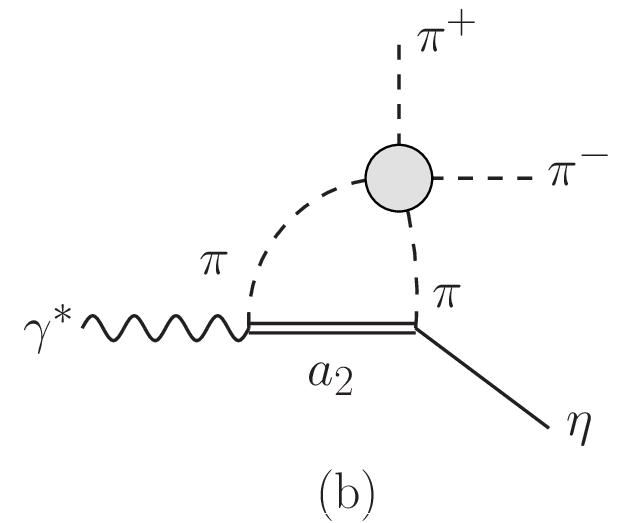}
   \caption{The reaction $e^+ e^- \to \eta \pi^+\pi^-$.
   The wiggly, dashed, solid, and double lines denote photons, pions, $\eta$, and $a_2$ meson, respectively.
   The gray blob stands for the $\pi\pi$ final-state interactions.} \label{fig:epem_pipieta}
\end{figure}

In this article, the program to pin down the $\eta$ transition form factor with high
accuracy will be extended to the doubly-virtual
form factor  $F_{\eta\gamma^*\gamma^*}(q^2,k^2)$ in the kinematic regime
$q^2\ll 1\GeV^2$ and $k^2>1\GeV^2$. 
The analysis is based on input from the reaction $e^+e^-\to\eta\pi^+\pi^-$, 
see Fig.~\ref{fig:epem_pipieta}, 
which plays a very similar role to data on $e^+e^-\to3\pi$ for the $\pi^0$ 
transition form factor~\cite{Hoferichter:2014vra,Hoferichter:2018dmo,Hoferichter:2018kwz}.
The method  provides access to the doubly-virtual form factor in a kinematic
regime where it is yet unknown.  In particular it allows for a test
whether the factorization ansatz 
\begin{equation}
\bar F_{\eta\gamma^*\gamma^*}\big(q^2,k^2\big)=\bar F_{\eta\gamma^*\gamma}\big(q^2\big)\bar F_{\eta\gamma\gamma^*}\big(k^2\big)
\label{eq:factor}
\end{equation}
for the normalized form factor $\bar F_{\eta\gamma^*\gamma^*}(q^2,k^2)= F_{\eta\gamma^*\gamma^*}(q^2,k^2)/F_{\eta\gamma^*\gamma^*}(0,0)$
is still valid in the kinematic regime specified above.\footnote{Strictly speaking, we test factorization 
for the isovector--isovector part of the transition form factor, as the isoscalar--isoscalar contribution
is small unless one of the virtualities hits a narrow isoscalar resonance, $q^2, k^2 \approx M_\omega^2$, $M_\phi^2$.} 
It should be stressed that although a direct measurement
of the doubly-virtual form factor via $e^+e^-\to\eta e^+e^-$ is in principle possible, we still expect 
our method to lead to higher accuracy, simply because the hadronic rates for $\gamma^*\to \eta\pi^+\pi^-$
are a factor $1/\alpha_{\rm em}^2$ larger than those for $\gamma^*\to\eta e^+e^-$. 
\esp 

\bsp
In Ref.~\cite{Kubis:2015sga} it was argued that the $a_2(1320)$,
a tensor resonance with $I^G (J^{PC}) = 1^- (2^{++})$,
should provide the leading left-hand-cut contribution to the decay amplitude for $\eta\to\pi^+\pi^-\gamma$, and
accordingly distort the spectra significantly, however, only beyond the kinematic region accessible
in the direct measurement of the decay. Interestingly all necessary parameters can be fixed from
data directly. The claim was corroborated by an analysis of the data
for $\eta'\to\pi^+\pi^-\gamma$~\cite{Hanhart:2016pcd,Ablikim:2017fll}.
In this work we therefore also include the $a_2$ contribution. 
Diagrammatically this amounts to including $a_2$ crossed-channel exchange as
shown in Fig.~\ref{fig:epem_pipieta}(b), 
in addition to the structureless vertex of Fig.~\ref{fig:epem_pipieta}(a).

Unfortunately, the differential data presently available for the reaction $e^+e^-\to\eta\pi^+\pi^-$ from the BaBar~\cite{Aubert:2007ef,TheBABAR:2018vvb}, SND~\cite{Achasov:2017kqm}, and CMD-3~\cite{Gribanov:2019qgw} collaborations are
rather limited: only BaBar and CMD-3 provide $\pi\pi$ spectra, and both with the shortcoming that they are not given at a fixed value of $k^2$, but for
$\sqrt{k^2}$ integrated in ranges from $1.0-4.5\GeV$~\cite{Aubert:2007ef}, $1.4-1.8\GeV$~\cite{TheBABAR:2018vvb}, or $1.3-1.8\GeV$~\cite{Gribanov:2019qgw}. 
As we argue below, analyzing the two BaBar data sets,
this integration limits the extraction accuracy of the form factor, since changes in the
$\pi\pi$ spectrum cannot only be induced by changes in the explicit dependence of the amplitude on the $\pi\pi$ invariant mass, but also by changes in the parameterization of the line shape of the $\rho'$.
The latter is the first excited resonance of the $\rho$ meson and
dominates the total cross section.

The paper is structured as follows. In Sect.~\ref{sec:disp} we generalize the
formalism of Ref.~\cite{Hanhart:2013vba} to the doubly-virtual $\eta$ transition form
factor, based on amplitudes for $\eta\to\pi^+\pi^-\gamma^*$.  Subsequently, we show how the dispersive representations for $\eta\to\pi^+\pi^-\gamma$
discussed in Refs.~\cite{Stollenwerk:2011zz,Kubis:2015sga} can be extended to the virtual-photon case of interest here.
In Sect.~\ref{sec:para} we discuss the parameterization used for virtualities beyond
$1\GeV^2$, which are beyond the range of applicability of the dispersive approach employed in this work. In Sect.~\ref{sec:results} we present and discuss the results
of our analysis of $e^+e^-\to\eta\pi^+\pi^-$ total and differential cross section data, without and with the inclusion of the $a_2$ contribution.  Section~\ref{sec:etaTFF} reflects on the impact of our findings for the $\eta$ transition form factor.  We close with a summary and discussion; some technicalities are relegated to an appendix.
\esp

\section{Dispersive representations}\label{sec:disp}
\subsection{Transition form factor of the $\eta$ meson}

The $\eta\to \gamma^*\gamma^*$ transition form factor (TFF)
$F_{\eta\gamma^*\gamma^*}(q^2,k^2)$ is defined by the matrix element 
\begin{align} 
    i \int \text{d}^4 x\  e^{iq \cdot x} & \braket{0 | T \lbrace j_\mu(x) j_\nu(0) \rbrace | \eta(q+k)} \notag\\
    &= \epsilon_{\mu \nu \alpha \beta} q^\alpha k^\beta F_{\eta\gamma^*\gamma^*}(q^2,k^2) \,,
\end{align}
where $j_\mu = (2\bar u\gamma_\mu u-\bar d\gamma_\mu d-\bar s\gamma_\mu s)/3$ denotes the electromagnetic current, 
$q$ and $k$ are the photon momenta, and the convention $\epsilon^{0123}=+1$ is used. The decay into two real photons is driven by the chiral anomaly~\cite{Wess:1971yu,Witten:1983tw}, which fixes the normalization $F_{\eta\gamma\gamma} \equiv F_{\eta\gamma^*\gamma^*}(0,0)$ of the TFF,
\begin{equation}
    \Gamma(\eta\to\gamma\gamma) = \frac{\pi \alpha_{\text{em}}^2 M_\eta^3}{4} \left| F_{\eta\gamma\gamma} \right|^2 \,,
\end{equation}
where $\alpha_{\text{em}} = e^2/(4\pi)$ is the fine-structure constant.

\bsp
Isospin symmetry allows us to decompose $F_{\eta\gamma^*\gamma^*}(q^2,k^2)$ into an isovector--isovector ($I=1$) and an isoscalar--isoscalar ($I=0$) part, which, for the normalized doubly-virtual TFF $\bar{F}_{\eta\gamma^*\gamma^*}(q^2,k^2)\equiv F_{\eta\gamma^*\gamma^*}(q^2,k^2)/F_{\eta\gamma\gamma}$, can be written according to
\begin{equation}
\bar{F}_{\eta\gamma^*\gamma^*}(q^2,k^2) \equiv 
 \bar F_{\eta\gamma^*\gamma^*}^{(I=1)}(q^2,k^2)+  \bar F_{\eta\gamma^*\gamma^*}^{(I=0)}(q^2,k^2) \,. 
 \label{eq:Fdef}
\end{equation}
As vector-meson-dominance estimates demonstrate, the isoscalar part is strongly suppressed~\cite{Hanhart:2013vba,Gan:2020aco}, such that for the TFF's low-energy properties, its contribution is within the isovector part's uncertainty.  
\esp

\begin{figure} 
    \centering
    \includegraphics[width=0.6\columnwidth]{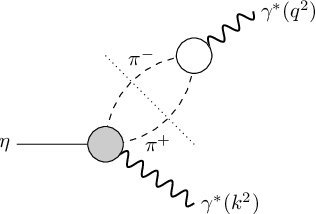}
    \caption{Diagrammatic representation of the dispersion relation for the isovector part of the $\eta$ TFF; see Eq.~\eqref{eq:DR}.
    }
    \label{Fig:etapipigamma_for_singlyvirtFF}
\end{figure}
A single-variable dispersion relation for the isovector part is at low energies dominated by the two-pion intermediate state, see Fig.~\ref{Fig:etapipigamma_for_singlyvirtFF}, which yields 
\begin{align}
\bar F_{\eta\gamma^*\gamma^*}^{(I=1)}(q^2,k^2) &= \frac{1}{96 \pi^{2} F_{\eta\gamma\gamma}} \notag\\
&\times \int_{4 M_\pi^2} ^\infty \diff t 
         \frac{t \sigma_\pi^3(t) F_{\eta\pi\pi\gamma^*}(t,k^2) [F_\pi^V(t)]^*}{t - q^2} \,,
    \label{eq:DR}
\end{align}
where $F_\pi^V(t)$ is the pion vector form factor, and $F_{\eta\pi\pi\gamma^*}(t,k^2)$ denotes the projection of the $\eta\to\pi^+\pi^-\gamma^*$ amplitude onto the pion--pion $P$-wave, to be defined formally below.  We expect the dispersion relation Eq.~\eqref{eq:DR} to hold up to $q^2 \sim 1\GeV^2$, even for a much wider range in the second argument $k^2$. 
We will briefly comment on the extent to which two pions alone saturate this isovector dispersion relation below in Sect.~\ref{sec:etaTFF}.

\subsection{Dipion amplitudes}
\bsp
The decay amplitude for $\eta(p_\eta) \to \pi^+(p_+)\pi^-(p_-)\gamma^{(*)}(k)$ can be written as
\begin{align}
\langle \pi^+(p_+)\pi^-(p_-) | & j_\mu(0) | \eta(p_\eta) \rangle \notag\\
&= \eps_{\mu\nu\alpha\beta} p_+^\nu p_-^\alpha k^\beta \, \F(s,t,u,k^2) \,, \label{eq:matrixelement}
\end{align}
with the Mandelstam variables given as  $s=(p_{\eta}-p_+)^2$, $t=(p_+ + p_-)^2$, and $u=(p_\eta-p_-)^2$, $s+t+u=\Sigma \equiv M_\eta^2+2M_\pi^2+k^2$.  
An expansion of $\F(s,t,u,k^2)$ in pion--pion partial waves proceeds in odd waves only, and is 
totally dominated by the $P$-wave~\cite{Kubis:2015sga}, denoted by $F_{\eta\pi\pi\gamma^*}(t,k^2)$ in the following,
\begin{equation}
F_{\eta\pi\pi\gamma^*}(t,k^2) = \frac{3}{4}\int_{-1}^1\diff z \big(1-z^2\big)\F(s,t,u,k^2) \,,
\end{equation}
where
\begin{equation}
z \equiv \cos\theta = \frac{s-u}{\sigma_\pi(t)\lambda^{1/2}(M_\eta^2,t,k^2)} \,, \quad \sigma_\pi(t) = \sqrt{1-\frac{4M_\pi^2}{t}} \,,
\end{equation}
and the Källén function is defined by 
\begin{equation}
    \lambda(a,b,c) = (a - b - c)^2 - 4bc \,.
\end{equation}

In the real-photon case $k^2=0$, the $\eta\to\pi^+\pi^-\gamma$ differential decay rate with respect to $t$ can be written as~\cite{Stollenwerk:2011zz}
\begin{align}
\frac{\diff\Gamma^{\eta}_{\pi\pi\gamma} }{\diff t} &= \left| F_{\eta\pi\pi\gamma^*}(t,0) \right|^2 \Gamma_0(t) \,, \notag\\
\Gamma_0(t) &= \frac{\alpha_{\text{em}} t \sigma_\pi^3(t)(M_\eta^2-t)^3}{1536\pi^2M_\eta^3} \,. \label{eq:main}
\end{align}
On the other hand, for $k^2 > (M_\eta+2M_\pi)^2$, the matrix element~\eqref{eq:matrixelement} can be assessed in the reaction 
$e^+e^-\to \eta\pi^+\pi^-$.  
We write the differential cross section as
\begin{equation}
        \label{Eq:diffxsec}
	    \frac{\diff \sigma}{\diff \sqrt{t}} (t, k^2) = N \left[ \frac{t}{4(k^2)^2} \lambda(t,k^2,M_{\eta}^2) \right]^{3/2} |F_{\eta\pi\pi\gamma^*}(t,k^2)|^2 \,,
\end{equation}
where the normalization is kept unspecified due to the BaBar differential cross sections being provided in arbitrary units~\cite{Aubert:2007ef,TheBABAR:2018vvb}. The total cross section data sets used, however, are provided in meaningful units. Hence, for the total cross section we use 
\begin{equation}
\sigma(k^2) = \frac{\alpha_{\text{em}}^2}{72 \pi}\int_{2M_\pi}^{\sqrt{k^2}-M_\eta} \diff \sqrt{t} \ \frac{1}{N} \frac{\diff \sigma}{\diff \sqrt{t}} (t, k^2) \,.
\end{equation}

A central element of the amplitude $F_{\eta\pi\pi\gamma^*}(t,k^2)$ is the universality of the elastic pion--pion $P$-wave final-state interactions as encoded in the Omn\`es function~\cite{Omnes:1958hv}
\begin{equation}
\Omega(t) = 
\exp\bigg\{\frac{t}{\pi}\int_{4M_\pi^2}^\infty\diff \tau \frac{\delta_1^1(\tau)}{\tau(\tau-t)}\bigg\} \,,
\end{equation}
where $\delta_1^1 (t)$ denotes the pion--pion $P$-wave phase shift.  In particular, $\Omega(t)$ contains the physics of the $\rho(770)$ resonance.  One of the most obvious applications to make use of the Omn\`es function is the pion vector form factor,
which can be extracted directly from data on $e^+e^-\to \pi^+\pi^-$ or $\tau^-\to\pi^-\pi^0\nu_\tau$. It can be written as
\begin{equation}
F_\pi^V(t) = R(t)\Omega(t) \,, 
\end{equation}
where $R(t)$ is a function free of cuts up to inelastic effects, which can in practice be approximated by a linear function up to $t \sim 1\GeV^2$~\cite{Hanhart:2013vba,Hanhart:2016pcd}. Utilizing the phase shift parameterization specified in~\ref{app:Pwave}
and fitting to $\tau$ decay data~\cite{Fujikawa:2008ma}, this linear function is found to be $R(t) = 1 + 0.126(2)\GeV^{-2} t$. 
Comparable parameterizations~\cite{Guo:2008nc,Hanhart:2012wi} were used in previous studies of the $\eta$ TFF~\cite{Stollenwerk:2011zz,Hanhart:2013vba}.
Similarly,  
\begin{equation}
    P(t,k^2) \equiv \frac{F_{\eta\pi\pi\gamma^*}(t,k^2)}{\Omega(t)}
    \label{eq:defP}
\end{equation}
is a function free of right-hand cuts in $t$ up to inelastic thresholds.  
In Refs.~\cite{Stollenwerk:2011zz,Babusci:2012ft} it was shown that for the real-photon case,
the function $P(t,0)$ is sufficiently well approximated as a linear polynomial
\begin{equation}
P^{(1)}(t,0)=A_{\pi\pi\gamma}^{\eta}(1+\alpha_\Omega t) \,, 
\label{eq:eqps}
\end{equation}
where $A_{\pi\pi\gamma}^{\eta}$ is a normalization constant,
to describe the high-accuracy $\eta\to \pi^+\pi^-\gamma$ decay data obtained by the KLOE collaboration~\cite{Babusci:2012ft}.
The slope parameter $\alpha_\Omega$ is found to be~\cite{Kubis:2015sga} 
\begin{equation}
 \alpha_\Omega = 1.52(6) \GeV^{-2} \,,
\label{eq:alphaval_om}
\end{equation}
with only the statistical uncertainty included.\footnote{The analyses~\cite{Stollenwerk:2011zz,Adlarson:2011xb,Babusci:2012ft} employ Eq.~\eqref{eq:defP} with $\Omega(t)$ replaced by $F_\pi^V(t)$, due to the observation that the linear slope in $R(t)$ relating both quantities is small.  The analogously defined parameter $\alpha$ was then determined to be $\alpha = 1.32(13) \GeV^{-2}$, combining statistical and systematic errors, which is slightly shifted compared to Eq.~\eqref{eq:alphaval_om}.}

As $P(t,0)$ is not expected to grow for large $t$ on fundamental grounds, the representation Eq.~\eqref{eq:eqps} is unlikely to hold beyond the small range in $t$ accessible directly
in the decay $\eta\to\pi^+\pi^-\gamma$.  Indeed, high-statistics data for the analogous decay $\eta'\to\pi^+\pi^-\gamma$ require a polynomial of \textit{second} order~\cite{Hanhart:2016pcd,Ablikim:2017fll} instead, motivating an alternative parameterization 
\begin{equation}
P^{(2)}(t,0)=A_{\pi\pi\gamma}^{\eta}\big(1+\alpha_\Omega t+\beta_\Omega t^2\big)  
\end{equation}
that we will also test below. 
A possible physical motivation for such a curvature term was given in Ref.~\cite{Kubis:2015sga}, where it was demonstrated that the left-hand cut induced by the $a_2(1320)$ distorts the linear behavior of $P(t,0)$ significantly beyond $t \leq M_\eta^2$.
In order to include the left-hand-cut contribution,
we need to generalize the dispersive representation of $F_{\eta\pi\pi\gamma^*}(t,0)$ according to~\cite{Kubis:2015sga}
\begin{equation}
F_{\eta\pi\pi\gamma^*}^{(a_2)} (t,0) = F_{a_2} (t,0) + \hat F_{a_2} (t,0) \,,
\label{eq:f1a2}
\end{equation}
with
\begin{align} 
F_{a_2} (t,0) &= \Omega (t) \bigg\{ A \bigl(1+ \alpha_\Omega[a_2] \, t \bigr) + \nonumber\\
& \qquad\qquad + \frac{t^2}{\pi} \int_{4 M_\pi^2}^\infty \frac{\diff \tau}{\tau^2} \frac{\sin \delta_1^1 (\tau) \hat F_{a_2} (\tau)}{|\Omega(\tau)| (\tau - t)} \bigg\} \,,  \nonumber \\
\hat F_{a_2} (t,0) &= \frac{3}{4} \int_{-1}^{1} \diff z \, \left(1 - z^2 \right) \F_{a_2} (s,t,u,0) \,, \label{eq:fa2}
\end{align}
where $\F_{a_2} (s,t,u,0)$ comprises the $a_2$ $s$- and $u$-channel exchange amplitudes (for real photons), see Ref.~\cite{Kubis:2015sga} for details.
The subtraction constant $\alpha_\Omega[a_2]$ that takes over the role of 
the slope parameter $\alpha_\Omega$ is shifted only marginally:
extracted from a fit to $\eta\to \pi^+\pi^-\gamma$ (where $t<0.25\GeV^2$),
it reads~\cite{Kubis:2015sga}
\begin{equation}
 \alpha_\Omega[a_2] = 1.42(6) \GeV^{-2} \,.
\label{eq:alphaval_om_a2}
\end{equation}

We define the function 
\begin{equation}
P^{(a_2)}(t,0) = \frac{F_{\eta\pi\pi\gamma^*}^{(a_2)} (t,0)}{\Omega(t)} 
\label{eq:poma2def}
\end{equation}
as a straightforward
generalization of Eq.~\eqref{eq:defP} (for $k^2=0$). This function is still free of right-hand cuts, but
now contains the left-hand cut due to $a_2$ exchange.
$P^{(a_2)}(t,0)$ can be approximated by
\begin{equation} \label{eq:Pa2}
P^{(a_2)}(t,0) \approx A_{\pi\pi\gamma}^\eta[a_2] \Big(1 + \big\{\alpha_\Omega[a_2]+\alpha_{a_2}\big\} t + \beta_{a_2}t^2 \Big)
\end{equation}
in the range $4M_\pi^2 \leq t \leq 1\GeV^2$. It turns out that
with $\alpha_{a_2}=0.28\GeV^{-2}$ and $\beta_{a_2}=-0.66\GeV^{-4}$,
this approximation works to better than 1\% accuracy.
The constant $A_{\pi\pi\gamma}^\eta[a_2]$ is chosen such
that $F_{\eta\pi\pi\gamma^*}(t,0)$ is normalized to the experimental rate for $\eta\to\pi^+\pi^-\gamma$.
Note that if the $a_2$ left-hand-cut contribution is turned off, the constants $A^\eta_{\pi \pi \gamma}$ and $A$, see Eq.~\eqref{eq:fa2}, coincide.
The overall strength of the $a_2$ contribution is phenomenologically known
from $a_2$ branching fractions to better than 10\% accuracy~\cite{Kubis:2015sga}.
Given the size of the errors in the data analyzed in the present study, we will neglect this 
source of uncertainty in what follows.

\bsp
The same method that allowed us to connect $\eta\to\pi^+\pi^-\gamma$ to the isovector component of
$\eta\to\gamma^*\gamma$
permits to connect $\gamma^*\to \eta\pi^+\pi^-$ to the isovector component $\gamma^*\to\eta\gamma^*$ and thus to the
doubly-virtual $\eta$ transition form factor. 
To  this end, we need to generalize the dispersion relation Eq.~\eqref{eq:DR}, and hence the description of $F_{\eta\pi\pi\gamma^*}(t,k^2)$, to an off-shell photon with invariant mass squared $k^2$.
We begin by noting that the left-hand-cut contribution of Eq.~\eqref{eq:f1a2} changes naturally for $k^2\neq 0$ due to the changed kinematics.  
We define
\begin{equation}
\label{Eq:inhom_omprob_sol}
F_{\eta\pi\pi\gamma^*}^{(a_2)}(t,k^2) = F_{a_2}(t,k^2) + \hat{F}_{a_2}(t,k^2) \,.
\end{equation}
The angular integral over the $a_2$ exchange amplitudes can be done analytically and yields the hat function (\textit{cf.}~\ref{app:hat} for details on the analytic continuation)
\begin{align}
\hat{F}_{a_2}(t,k^2)
&=\frac{8 c_T g_T}{\sqrt{3} F_\pi^3} \left(\frac{M_{\eta}^2-M_\pi^2}{M_{a_2}^2} - 1 + \Bigg[M_{a_2}^2 - \Sigma +2t \right.  \notag\\
& \left.  +\frac{(k^2-M_\pi^2)(M_{\eta}^2-M_\pi^2)}{M_{a_2}^2}\Bigg] \frac{3Q(y(t,k^2))}{2M_{a_2}^2-\Sigma+t}	\right) \,, \notag \\
Q(y)&=y\left[\frac{1}{2}(1-y^2)\log\left(\frac{y+1}{y-1}\right)  + y\right] \,, \notag \\
y(t,k^2)&=\frac{2M_{a_2}^2-\Sigma+t}{\sigma_\pi(t) \sqrt{\lambda(t,k^2,M_\eta^2)}} \,,  \label{Eq:hat_function}
\end{align}
where the coupling constant $g_T$ ($c_T$) has been determined from the branching ratio of $a_2 \to \pi \eta$ ($a_2 \to \pi \gamma$)~\cite{Kubis:2015sga} and $F_\pi=92.28(10)\MeV$~\cite{Zyla:2020zbs} is the pion decay constant. 
The dispersion integral contained in
\begin{align}
F_{a_2}(t,k^2) &= \Omega (t) \bigg\{ A \bigl(1+ \alpha^*_\Omega[a_2](k^2) \ t \bigr) \notag \\
& \qquad\qquad + \frac{t^2}{\pi} \int_{4 M_\pi^2}^\infty \frac{\diff \tau}{\tau^2} \frac{\sin \delta_1^1 (\tau) \hat F_{a_2} (\tau,k^2)}{|\Omega(\tau)| (\tau - t)} \bigg\}
\label{Eq:Fa2ksq}
\end{align}
can be carried out numerically via path deformation into the complex plane in order to circumvent the singularities of the integrand, inspired by the methods of Ref.~\cite{Gasser:2018qtg}. Due to the need to deform the integration path into the complex plane, we utilize a $\pi \pi$ phase shift $\delta_1^1$ calculated via the inverse-amplitude method (we employ an approximation to the two-loop representation of Ref.~\cite{Niehus:2020gmf}), which provides us with an analytic expression; see~\ref{app:Pwave}.  Equations~\eqref{Eq:hat_function} and \eqref{Eq:Fa2ksq} encode a nontrivial entanglement of the dependences on $t$ and $k^2$ that clearly does not factorize in a simple manner.

We now make the following ansatz for the complete $\gamma^* \to \eta \pi^+ \pi^-$ form factor:
\begin{equation}
    F_{\eta\pi\pi\gamma^*}(t,k^2) = F_{\eta\pi\pi\gamma^*}^{(a_2)}(t,k^2) \tilde{F}_{\eta\gamma\gamma^*}(k^2) \,, \label{eq:semifactorization}
\end{equation}
\textit{i.e.}, any additional $k^2$ dependence is assumed to be multiplicative (or factorizing).  
Note that $\tilde{F}_{\eta\gamma\gamma^*}(k^2)$ cannot be treated within the formalism
of Ref.~\cite{Hanhart:2013vba} for an application to $e^+e^-\to\eta\pi^+\pi^-$ data, since with $k^2>1\GeV^2$ it is to be evaluated in a kinematic regime
where the original method, relying on elastic unitarity, is no longer applicable. However, as will be shown in Sect.~\ref{sec:para},
for the analysis aimed at here we only need a convenient parameterization of $\tilde{F}_{\eta\gamma\gamma^*}(k^2)$. 
Alternatively, we will revert to the simpler approach to replace $F_{\eta\pi\pi\gamma^*}^{(a_2)}(t,k^2)$ by $F_{\eta\pi\pi\gamma^*}(t,0)$, given by either the linear or the quadratic approximations $P^{(1,2)}(t,0)$.  In these limits, $F_{\eta\pi\pi\gamma^*}^{(a_2)}(t,k^2)$ and, via the dispersion integral~\eqref{eq:DR}, the (isovector part of the) doubly-virtual TFF, become factorizing in their dependences on the two respective variables, and we can check the consistency of the assumption by testing to which extent the linear subtraction constants of Eqs.~\eqref{eq:fa2} and \eqref{Eq:Fa2ksq} fulfill $\alpha_\Omega[a_2] = \alpha^*_\Omega[a_2]$, \textit{i.e.}, to which extent $\alpha^*_\Omega[a_2](k^2)$ can (on average) be approximated by a constant consistent with the value for the real-photon case. 
In this limit (only), $\tilde{F}_{\eta\gamma\gamma^*}(k^2)$ also corresponds to the normalized singly-virtual $\eta$ TFF $\bar{F}_{\eta\gamma\gamma^*}(k^2)$.  
\esp
 
In order to fix $\alpha^*_\Omega[a_2](k^2)$ and $\tilde F_{\eta\gamma\gamma^*}(k^2)$, which
provide the input for the calculation of  the doubly-virtual transition form factor $F_{\eta\gamma^*\gamma^*}(q^2,k^2)$, data on $e^+e^-\to\eta\pi^+\pi^-$ need to be analyzed, with particular attention to the distributions with the dipion invariant masses. 
However, data for the $\pi\pi$ spectrum are available only with a simultaneous integration
over the initial energy $k^2$. Therefore it is not possible to extract values
for $\alpha^*_\Omega[a_2](k^2)$ as a function of $k^2$, but only to extract some averaged value
$\bar\alpha^*_\Omega[a_2]$. Moreover, since the value of $\bar\alpha^*_\Omega[a_2]$ also influences the
shape of the total cross section for $e^+e^-\to \eta\pi^+\pi^-$, a combined fit
to the $\pi\pi$ spectrum and the total cross section is mandatory,
which calls for a parameterization of the data in kinematic regimes that cannot be 
captured by the dispersion
integrals employed in this work. We describe this parameterization in the following section.
\esp

\begin{figure} 
    \centering
    \includegraphics[width=\linewidth]{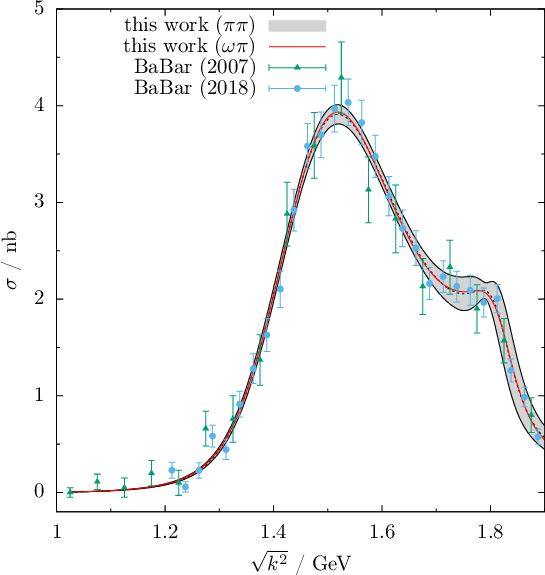}
    \caption{Fit to the total cross section from BaBar 2007~\cite{Aubert:2007ef} and 2018~\cite{TheBABAR:2018vvb} up to $\sqrt{k^2}=\SI{1.9}{\GeV}$ for $s_c=\SI{3}{\GeV^2}$ (see Eq.~\eqref{Eq:fullmodel_highenergy} for details). The gray band around the central black-dashed curve represents the full model with $\pi \pi$ widths (\textit{cf.}\ Eq.~\eqref{Eq:pipi-width}) used in the Breit--Wigner parameterizations, while the red line corresponds to the central value of the full model with $\omega \pi$ widths (\textit{cf.}\ Eq.~\eqref{eq:width-piomega}) for $\rho'$ and $\rho''$. The error band of this curve is very close to the gray one. The central values of both variants display only tiny differences between each other.
    }
    \label{Fig:totxsec}
\end{figure}

\section{Form factor parameterization above $1\GeV$}
\label{sec:para}
\subsection{Inclusion of excited isovector vector resonances}

\bsp
For the fit to the total cross section we need to parameterize the function $\tilde{F}_{\eta\gamma\gamma^*}(k^2)$
for $k^2>1\GeV^2$.
As one can see from Fig.~\ref{Fig:totxsec}, the total cross section for 
$e^+e^-\to\eta\pi^+\pi^-$ shows a very prominent contribution from an excited $\rho$ resonance, the $\rho'$ or $\rho(1450)$.
It turns out that the low-energy side of the spectrum calls for
an inclusion of the $\rho(770)$ resonance, which enters through its upper tail.
Moreover, 
the addition of a higher resonance, the $\rho''$ or $\rho(1700)$, improves the data fit significantly. 
Note that most experimental or phenomenological analyses of the reaction $e^+e^-\to\eta\pi^+\pi^-$ employ two or three vector resonances~\cite{Akhmetshin:2000wv,Dai:2013joa,Volkov:2013zba,Qin:2020udp}. 
It should be stressed that what is crucial for the analysis is not that we have the correct physics optimally
built into the parameterization of the function $\tilde{F}_{\eta\gamma\gamma^*}(k^2)$, but all we need is a convenient representation
that describes the data for the quantity of interest for this analysis: the $k^2$-integrated 
$\pi\pi$ spectrum. Since $\tilde{F}_{\eta\gamma\gamma^*}(k^2)$ enters here as a weight factor---{\it cf.}\ Eq.~\eqref{Eq:diffxsec}---it is sufficient to parameterize
$\tilde{F}_{\eta\gamma\gamma^*}(k^2)$ by a sum of Breit--Wigner functions. Clearly in doing so we should not
expect the resonance parameters to agree with those of the Particle Data Group (PDG), since Breit--Wigner
parameters are reaction dependent and a sum of Breit--Wigner functions violates unitarity.
More explicitly, the Breit--Wigner (BW) functions we employ are given by
\begin{equation}
    BW_j(k^2) = \frac{1}{M_j^2 - k^2 - i\sqrt{k^2}\Gamma_j(k^2)} \,, \label{eq:defBW}
\end{equation}
with $j \in \lbrace \rho,\rho',\rho'' \rbrace$ and the energy-dependent widths written as~\cite{Achasov:1984ru}
\begin{equation}
\label{Eq:pipi-width}
    \Gamma_j(k^2) = \Gamma_j(M_j^2) \frac{M_j^2}{k^2} \frac{q_{\pi\pi}^3(k^2)}{q_{\pi\pi}^3(M_j^2)}  \theta\big(k^2-4M_\pi^2\big)\,,
\end{equation}
where the center-of-mass momentum is given by
\begin{equation}
q_{ab}(s) = \frac{\sqrt{\lambda(s,M_a^2,M_b^2)}}{2\sqrt{s}} \,. 
\end{equation}
Additionally, we employ another variant, in which the $\rho'$ and $\rho''$ widths are parameterized by the decay into the $\omega \pi$ final state: 
\begin{equation}
    \Gamma_j(k^2)= \Gamma_j(M_j^2) 
    \frac{q_{\omega\pi}^3(k^2)}{q_{\omega\pi}^3(M_j^2)}
    \theta\big(k^2 - (M_\omega + M_\pi)^2\big) \,. \label{eq:width-piomega}
\end{equation}

We parameterize the function $\tilde{F}_{\eta\gamma\gamma^*}(k^2) \propto \sum_{j} c_j M_j^2 BW_j(k^2)$, where $j \in \lbrace \rho,\rho',\rho'' \rbrace$. Through the normalization constraint $\tilde{F}_{\eta\gamma\gamma^*}(0)=1$ we eliminate the coupling $c_{\rho}$:
\begin{align}
    \label{Eq:BWsum}
    &\tilde{F}_{\eta\gamma\gamma^*}(k^2) = \notag \\ & 
    \frac{M_\rho^2 BW_\rho(k^2) + c_{\rho'} M_{\rho'}^2 BW_{\rho'}(k^2) + c_{\rho''} M_{\rho''}^2 BW_{\rho''}(k^2)}{1+c_{\rho'}+c_{\rho''}} \,.
\end{align}
The remaining couplings are used as fit parameters.  While the (large) Breit--Wigner coupling to the $\rho'$ is taken to be real as required by unitarity, we see a substantial improvement in the fit quality if the (small) coupling to the $\rho''$ is allowed to take complex values, which likely subsumes the effects of yet higher resonances and the more complicated coupled-channel structure in that energy region. 
Additionally, the masses $M_{\rho '}$ and $M_{\rho''}$ as well as the widths $\Gamma_{\rho'}(M_{\rho'}^2)$ and $\Gamma_{\rho''}(M_{\rho''}^2)$ of the $\rho'$ and $\rho''$ are fitted to the data. For the $\rho$ we use the PDG values for mass and width.
\esp

Since we do not expect the form factor ansatz in Eq.~\eqref{Eq:inhom_omprob_sol} to hold up to arbitrarily high energies, we employ the following prescription for a certain cutoff value $s_c$:
\begin{align}
\label{Eq:fullmodel_highenergy}
	&F_{\eta\pi\pi\gamma^*}(t,k^2) =  \\
	&\begin{cases}
		\Omega(t) \left( L(t) + \dfrac{t^2}{\pi} D(t,k^2)\right)  + \hat{F}_{a_2}(t,k^2) \quad \text{for} \quad t,k^2\leq s_c \,,\\
		\Omega(t) \left( L(s_c) + \dfrac{s_c^2}{\pi} D(s_c,k^2)\right) + \hat{F}_{a_2}(s_c,k^2) \dfrac{s_c}{t}
		\\  \hfill \text{for} \quad t>s_c ~\land~ k^2 \leq s_c \,, \\
		\Omega(t) \left( L(t) + \dfrac{t^2}{\pi} D(t,s_c)\dfrac{s_c}{k^2}\right) + \hat{F}_{a_2}(t,s_c) \dfrac{s_c}{k^2} \\ \hfill \text{for} \quad t \leq s_c ~\land~ k^2 > s_c \,,\\
		\Omega(t) \left( L(s_c) + \dfrac{s_c^2}{\pi} D(s_c,s_c) \dfrac{s_c}{k^2}\right) +
		\hat{F}_{a_2}(s_c,s_c) K(t,k^2) \\ \hfill \text{for} \quad t,k^2>s_c \,, 
	\end{cases} \notag
\end{align}
with
\begin{align}
    L(t) &= A \bigl(1+ \alpha^*_\Omega[a_2] \, t \bigr) \,, \notag \\
    K(t,k^2) &= \frac{s_c(t+s_c)(k^2 + s_c)}{2tk^2(t+k^2)} \,,
\end{align}
and the dispersion integral
\begin{equation}
    D(t,k^2) = \int_{4 M_\pi^2}^\infty \frac{\diff \tau}{\tau^2} \frac{\sin \delta_1^1 (\tau) \hat F_{a_2} (\tau,k^2)}{|\Omega(\tau)| (\tau - t)} \,.
\end{equation}
This continuation ensures that the form factor drops off like $1/t$ and $1/k^2$ for high energies and is continuous everywhere when going from one region to another. Through treating the cutoffs in $t$ and in $k^2$ in the same way, it reflects a symmetry that is important for a later application in the $\eta$ transition form factor in the HLbL contribution to the muon $g-2$. We test the independence of the choice of continuation point $s_c$ by varying it in the fits according to $s_c \in \lbrace 3,\,4,\,5 \rbrace \,\GeV^2$.

\subsection{Finite-width effects in the $a_2$ exchange}
\label{Sec:a2}
\bsp
In the tensor model used to describe the $a_2$-exchange diagrams, the $a_2$ meson is assumed to have no width. However, in reality the $a_2(1320)$ has a width of about \SI{100}{\MeV}, and in the reaction $e^+e^-\to\eta\pi^+\pi^-$, where the $a_2$ can be produced quasi on-shell, this effect is surely not negligible. To cure this caveat, 
we define a dispersively improved Breit--Wigner function $\overline{BW}(s)$, using the imaginary part of $BW(s)$ as in Eq.~\eqref{eq:defBW} as the spectral function above a certain threshold
$s_\text{thr}$~\cite{Lomon:2012pn,Moussallam:2013una},
\begin{align}
		\overline{BW}(s) &= \frac{1}{\pi}\int_{s_\mathrm{thr}}^{\infty} \diff x\ \frac{\operatorname{Im} BW(x)}{\textcolor{black}{x-s}} \notag\\
		&= \frac{1}{\pi} \int_{s_\mathrm{thr}}^{\infty} \diff x \ \operatorname{Im} (BW(x)) \cdot \textcolor{black}{BW(s;x;\Gamma=0)} \,,
\end{align}
where in the last step we have written the Cauchy kernel as a Breit--Wigner function with mass parameter $\sqrt{x}$ and width zero.  
We therefore generalize both dispersion integral $D(t,k^2)$ and hat function $\hat{F}_{a_2}(t,k^2)$ with the $a_2$ mass dispersed according to this spectral function,
\begin{equation}
\label{Eq:MA2var}
	\overline{C}(t,k^2) = \frac{1}{\pi} \int_{s_\mathrm{thr}}^{\infty} \textcolor{black}{\diff M_{a_2}^2} \ \operatorname{Im} (BW(M_{a_2}^2)) \cdot \textcolor{black}{C(t,k^2;M_{a_2}^2)} \,,
\end{equation}
where $C \in \lbrace D, \hat{F}_{a_2} \rbrace$.
We employ the parameterization~\cite{VonHippel:1972fg,Beladidze:1993km}
\begin{align}
\Gamma(s) &= \Gamma_{a_2} \sum\limits_{i \in \lbrace \eta, \rho \rbrace}  p_i \frac{M_{a_2}^2}{s} \frac{q_{i\pi}(s)}{q_{i\pi}(M_{a_2}^2)} \frac{T(q_{i\pi}(s)R)}{T(q_{i\pi}(M_{a_2}^2)R)} \notag\\ & \qquad\qquad \quad \times \theta\big(s-(M_i+M_\pi)^2\big)\,, \notag\\
\label{eq:width-a2}
T(x)&=\frac{x^4}{9 + 3x^2 +x^4} \,,
\end{align}
where $M_\rho$ ($M_\eta$) is the mass of the $\rho$ ($\eta$) meson, $\Gamma_{a_2}$ the total $a_2$ width, and the barrier factor is given by $R = \SI{5.2}{\GeV^{-1}}$~\cite{Beladidze:1993km}. This energy-dependent width explicitly takes into account the $a_2$ decays into $\pi \eta$ and $\pi \rho$ with branching ratios $p_\eta = \num{0.17}$ and $p_\rho = \num{0.83}$.
\esp

We evaluate $C(t,k^2;M_{a_2}^2)$ for \num{20} different values of $M_{a_2}^2$ in the range $\left[s_\text{thr},\SI{2.3}{\GeV^2}\right]$, where $s_\text{thr} = (M_\eta + M_\pi)^2$. For $D$ we use a grid of $\num{107} \times \num{150}$ values for $t,k^2 \in [4M_\pi^2,\SI{3}{\GeV^2}]$. The integral of Eq.~(\ref{Eq:MA2var}) was then carried out using the trapezoidal rule for each point of the $(t,k^2)$ grid. In the fitting routine we make the replacements
\begin{equation}
\label{Eq:a2width_repl}
	\hat{F}_{a_2}(t,k^2) \to \hat{\overline{F}}_{a_2}(t,k^2) \quad \text{and} \quad D(t,k^2) \to \overline{D}(t,k^2) \,.
\end{equation}

\section{Data analysis of $e^+e^-\to \eta\pi^+\pi^-$} \label{sec:results}

\bsp
In the following, we analyze the $e^+e^-\to \eta\pi^+\pi^-$ data by the BaBar collaboration~\cite{Aubert:2007ef,TheBABAR:2018vvb} in three successively refined models for the $\pi\pi$ invariant mass distribution: 
assuming a linear or a quadratic polynomial multiplying the final-state interactions as parameterized by the Omn\`es function, and using the full-fledged description of left-hand cuts due to the $a_2$, which automatically generates breaking of the factorization assumption.
We fit to data from BaBar published in 2007~\cite{Aubert:2007ef} and 2018~\cite{TheBABAR:2018vvb}. The differential cross section data from 2007 are integrated over $\sqrt{k^2}$ in  $[1.0,\, 4.5]\,\si{\GeV}$, the ones from 2018 in the range $\sqrt{k^2} \in [1.4,\, 1.8]\,\si{\GeV}$. The $\pi\pi$ spectrum is dominated by the $\rho$, we choose to fit the differential cross section data up to $\sqrt{t}=\SI{1}{\GeV}$. The tail of the Breit--Wigner parameterization of $\tilde{F}_{\eta\gamma\gamma^*}(k^2)$ is rather uncontrolled for larger $k^2$, thus we fit the total cross section data up to $\sqrt{k^2}=\SI{1.9}{\GeV}$ only. 

\subsection{Linear model}
As in the first dispersive attempts that dealt with the $\eta$ TFF~\cite{Hanhart:2013vba,Stollenwerk:2011zz} we employ a linear function as part of $F_{\eta\pi\pi\gamma^*}(t,k^2)$, which does not include any effects of the $a_2$-exchange left-hand-cut contribution: 
\begin{equation}
    F_{\eta\pi\pi\gamma^*}^\text{lin.} (t,k^2) = \Omega(t) A (1 + \alpha_\Omega^* t) \tilde F_{\eta\gamma\gamma^*}(k^2) \,. \label{Eq:ModelLinear}
\end{equation}
For values of the $\pi \pi$ invariant mass squared larger than a certain cutoff parameter $s_p$ we let the form factor fall off like $1/t$ by setting the linear polynomial to a constant. The constant term in the polynomial part of Eq.~\eqref{Eq:ModelLinear} is fixed by fitting the corresponding linear representation to $\eta \to \pi^+ \pi^- \gamma$ data measured by KLOE~\cite{Babusci:2012ft} to $A=\SI{18.0 \pm 0.4}{\GeV^{-3}}$. The results of these fits with cutoff parameter $\sqrt{s_p} \in \lbrace 0.9,\, 1.0,\, 1.2 \rbrace\GeV$ are shown in Table~\ref{Tab:FitresLinPolynomial}. The central finding is that the extracted values for $\alpha_\Omega^*$ are consistently and significantly smaller than those found for $\alpha_\Omega$, see Eq.~\eqref{eq:alphaval_om}.

\begin{table*} 
\renewcommand{\arraystretch}{1.3}
	\centering
	\caption{
		Results for the simultaneous fit to differential and total cross section data from BaBar (2007)~\cite{Aubert:2007ef} and BaBar (2018)~\cite{TheBABAR:2018vvb}, employing the linear model of Eq.~\eqref{Eq:ModelLinear}. The differential cross section data have been fitted up to \SI{1}{\GeV}, the total cross section data up to \SI{1.9}{\GeV}. $(\chi^2/\text{dof})_\text{c}^\text{y}$ labels the reduced $\chi^2$ for the respective data sets, $c=$dif/tot for the differential/total cross section of BaBar in year y. The parameter $N_\text{y}$ labels the unphysical normalization of the differential cross section data of BaBar in year y.  Columns 2--4 refer to fits employing the BW parameterization of $\rho'$ and $\rho''$ widths according to $\pi\pi$ decays, columns 5--7 display those relying on the $\omega\pi$ running width model.
	}
	\label{Tab:FitresLinPolynomial}
	\begin{tabular}{l c c c c c c}
	\toprule
	$\rho',\rho''$ width & \multicolumn{3}{c}{$\pi\pi$} & \multicolumn{3}{c}{$\omega\pi$} 
	\\
	$\sqrt{s_p}$/$\si{\GeV}$ & \num{0.9} & \num{1.0} & \num{1.2} & \num{0.9} & \num{1.0} & \num{1.2}\\ \midrule
	$(\chi^2/\text{dof})_\text{dif}^{07}$ & \num{2.14} & \num{2.09} & \num{2.09}& \num{1.97} & \num{1.94} & \num{1.94} \\
	$(\chi^2/\text{dof})_\text{dif}^{18}$ & \num{2.00} & \num{1.93} & \num{1.93}& \num{2.00} & \num{1.93} & \num{1.93}\\
	$(\chi^2/\text{dof})_\text{tot}^{07+18}$ & \num{0.86} & \num{0.85} & \num{0.85} & \num{0.83} & \num{0.83} & \num{0.83}\\
	$\bar{\alpha}^*_\Omega/\si{GeV^{-2}}$ & \num{0.74 \pm 0.14} & \num{0.79 \pm 0.15} & \num{0.79 \pm 0.15} & \num{0.73 \pm 0.14} & \num{0.76 \pm 0.14} & \num{0.76 \pm 0.14} \\
	$M_{\rho'}/\si{\MeV}$ & \num{1517.3 \pm 6.8} & \num{1517.2 \pm 6.8} & \num{1517.2 \pm 6.8} & \num{1632 \pm 18} & \num{1629 \pm 21} & \num{1629 \pm 21}\\
	$\Gamma_{\rho'}(M_{\rho'}^2)/\si{\MeV}$ & \num{365 \pm 29} & \num{363 \pm 29} & \num{363 \pm 29} & \num{680 \pm 110} & \num{667 \pm 99} & \num{667 \pm 99}\\
	$c_{\rho'}/10^{-1}$ & \num{-2.62 \pm 0.19} & \num{-2.58 \pm 0.18} & \num{-2.58 \pm 0.18}& \num{-2.71 \pm 0.20} & \num{-2.67 \pm 0.22} & \num{-2.67 \pm 0.22}\\
	$c_{\rho''}/10^{-2}$ & \num{1.04 \pm 0.44} & \num{1.04 \pm 0.44} & \num{1.03 \pm 0.44} & \num{1.16 \pm 0.46} & \num{1.16 \pm 0.45} & \num{1.17 \pm 0.46}\\
	$\phi_{\rho''}/\si{\radian}$ & \num{4.53 \pm 0.28} & \num{4.53 \pm 0.28} & \num{4.54 \pm 0.28}& \num{2.51 \pm 0.26} & \num{2.49 \pm 0.27} & \num{2.48 \pm 0.27}\\
	$M_{\rho''}/\si{\MeV}$ & \num{1816 \pm 10} & \num{1816 \pm 10} & \num{1816 \pm 10} & \num{1826 \pm 13} & \num{1827 \pm 13} & \num{1827 \pm 13}\\
	$\Gamma_{\rho''}(M_{\rho''}^2)/\si{\MeV}$ & \num{109 \pm 30} & \num{110 \pm 30} & \num{110 \pm 30} & \num{126 \pm 31} & \num{128 \pm 34} & \num{128 \pm 34}\\
	$N_{07}/10^1\,\si{\text{arb. units}}$ & \num{4.97 \pm 0.18} & \num{5.00 \pm 0.18} & \num{5.01 \pm 0.18} & \num{4.61 \pm 0.18} & \num{4.62 \pm 0.20} & \num{5.63 \pm 0.18}\\
	$N_{18}/10^1\,\si{\text{arb. units}}$ & \num{9.84 \pm 0.21} & \num{9.86 \pm 0.21} & \num{9.86 \pm 0.21} & \num{10.55 \pm 0.21} & \num{9.86 \pm 0.21} & \num{9.86 \pm 0.21}\\
	\bottomrule
	\end{tabular}
\renewcommand{\arraystretch}{1.0}
\end{table*}
\subsection{Quadratic model}
\begin{table*} 
\renewcommand{\arraystretch}{1.3}
	\centering
	\caption{
	Fit results as in Table~\ref{Tab:FitresLinPolynomial}, but
	employing the quadratic model approximation of Eq.~\eqref{Eq:ModelPoly}. 
	}
	\label{Tab:FitresPolynomial}
	\begin{tabular}{l c c c c c c}
	\toprule
	$\rho',\rho''$ width & \multicolumn{3}{c}{$\pi\pi$} & \multicolumn{3}{c}{$\omega\pi$}\\
	$\sqrt{s_p}$/$\si{\GeV}$ & \num{0.9} & \num{1.0} & \num{1.2} & \num{0.9} & \num{1.0} & \num{1.2}\\ \midrule
	$(\chi^2/\text{dof})_\text{dif}^{07}$ & \num{2.31} & \num{2.27} & \num{2.27} & \num{2.07} & \num{2.04} & \num{2.03} \\
	$(\chi^2/\text{dof})_\text{dif}^{18}$ & \num{2.17} & \num{2.12} & \num{2.12} & \num{2.17} & \num{2.12} & \num{2.12}\\
	$(\chi^2/\text{dof})_\text{tot}^{07+18}$ & \num{0.84} & \num{0.83} & \num{0.83} & \num{0.82} & \num{0.82} & \num{0.82}\\
	$\bar{\alpha}^*_\Omega[a_2]/\si{GeV^{-2}}$ & \num{1.43 \pm 0.19} & \num{1.52 \pm 0.20} & \num{1.52 \pm 0.20} & \num{1.38 \pm 0.17} & \num{1.44 \pm 0.17} & \num{1.44 \pm 0.17} \\
	$M_{\rho'}/\si{\MeV}$ & \num{1527.2 \pm 8.5} & \num{1527.5 \pm 8.8} & \num{1527.5 \pm 8.9}  & \num{1632 \pm 21} & \num{1630 \pm 21} & \num{1629 \pm 20}\\
	$\Gamma_{\rho'}(M_{\rho'}^2)/\si{\MeV}$ & \num{363 \pm 36} & \num{360 \pm 20} & \num{360 \pm 36} & \num{650 \pm 100} & \num{640 \pm 100} & \num{640 \pm 100}\\
	$c_{\rho'}/10^{-1}$ & \num{-2.14 \pm 0.20} & \num{-2.07 \pm 0.20} & \num{-2.07 \pm 0.20} & \num{-2.21 \pm 0.21} & \num{-2.15 \pm 0.21} & \num{-2.16 \pm 0.20}\\
	$c_{\rho''}/10^{-2}$ & \num{1.24 \pm 0.57} & \num{1.28 \pm 0.60} & \num{1.27 \pm 0.60}  & \num{1.15 \pm 0.45} & \num{1.17 \pm 0.47} & \num{1.17 \pm 0.46}\\
	$\phi_{\rho''}/\si{\radian}$ & \num{4.35 \pm 0.38} & \num{4.35 \pm 0.41} & \num{4.35 \pm 0.41}  & \num{2.57 \pm 0.27} & \num{2.55 \pm 0.28} & \num{2.55 \pm 0.28}\\
	$M_{\rho''}/\si{\MeV}$ & \num{1814 \pm 12} & \num{1814 \pm 12} & \num{1815 \pm 13} & \num{1829 \pm 14} & \num{1830 \pm 15} & \num{1830 \pm 15}\\
	$\Gamma_{\rho''}(M_{\rho''}^2)/\si{\MeV}$ & \num{130 \pm 33} & \num{133 \pm 34} & \num{133 \pm 34} & \num{136 \pm 36} & \num{140 \pm 37} & \num{140 \pm 36}\\
	$N_{07}/10^1\,\si{\text{arb. units}}$ & \num{5.05 \pm 0.18} & \num{5.08 \pm 0.18} & \num{5.08 \pm 0.18} & \num{4.48 \pm 0.20} & \num{4.48 \pm 0.20} & \num{4.48 \pm 0.20}\\
	$N_{18}/10^1\,\si{\text{arb. units}}$ & \num{9.79 \pm 0.21} & \num{9.80 \pm 0.21} & \num{9.80 \pm 0.21} & \num{9.79 \pm 0.21} & \num{9.80 \pm 0.21} & \num{9.80 \pm 0.20}\\
	\bottomrule
	\end{tabular}
\renewcommand{\arraystretch}{1.0}
\end{table*}

As noted in Ref.~\cite{Kubis:2015sga} for the real-photon case, the inclusion of the $a_2$-induced left-hand-cut contribution adds a curvature part to the form factor $F_{\eta\pi\pi\gamma^*}(t,k^2)$. In the quadratic model, we do not use the explicit form of the $a_2$ exchange amplitudes, but restrict ourselves to the approximation in Eq.~\eqref{eq:Pa2}. The fit of this representation to $\eta \to \pi^+ \pi^- \gamma$ decay data yields $A=\SI{17.4 \pm 0.4}{\GeV^{-3}}$. Note that this form omits all the $k^2$-dependence in $F_{\eta\pi\pi\gamma^*}(t,k^2)$ other than through the multiplicative factor $\tilde F_{\eta\gamma\gamma^*}(k^2)$, \textit{cf.}\ Eq.~\eqref{Eq:BWsum}:
\begin{equation}
\label{Eq:ModelPoly}
    F_{\eta\pi\pi\gamma^*}^\text{quad.}(t,k^2) = \Omega(t)AP_\Omega[a_2](t) \tilde F_{\eta\gamma\gamma^*}(k^2) \,,
\end{equation}
where the linear coefficient $\bar{\alpha}^*_\Omega[a_2]$ is used as a fit parameter and the parameters $\alpha_{a_2}$ and $\beta_{a_2}$ are held fixed at their respective values given below Eq.~\eqref{eq:Pa2}. For values of the $\pi \pi$ invariant mass squared larger than a certain cutoff parameter $s_p$ we, again, let the form factor fall off like $1/t$ by the prescription
\begin{equation}
\label{Eq:QuadCont}
    F_{\eta\pi\pi\gamma^*}^\text{quad.}(t,k^2) = \Omega(t)AP_\Omega[a_2](s_p) \tilde F_{\eta\gamma\gamma^*}(k^2)  \quad \text{for} \quad t>s_p \,.
\end{equation}
The fit results for cutoff parameter $\sqrt{s_p} \in \lbrace 0.9,\, 1.0,\, 1.2 \rbrace\GeV$ are displayed in Table~\ref{Tab:FitresPolynomial}. 
Compared to the results of the linear model, the linear parameter $\bar{\alpha}^*_\Omega[a_2]$ comes out substantially closer to the value of the linear subtraction constant $\alpha_\Omega[a_2]$ for all tested values of $s_p$, \textit{cf.}\ Eq.~\eqref{eq:alphaval_om_a2}, which has been obtained from $\eta \to \pi^+\pi^- \gamma$ decay data, and compatible within uncertainties throughout. The sizable shift of the slope between the linear and quadratic model can be understood in the following manner: in the real-photon case, the main impact of the $a_2$ contribution on the $\pi \pi$ spectrum is to provide a downward curvature, with the quadratic coefficient $\beta_{a_2}$ being sizable and negative. Accordingly, a fit with a linear term only must find a reduced strength in order to reproduce the data in the region of the $\rho$ resonance.

\subsection{Full model}

\begin{table*} 
\renewcommand{\arraystretch}{1.3}
	\centering
	\caption{
	Fit results as in Table~\ref{Tab:FitresLinPolynomial}, but employing the full model.
	}
	\label{Tab:FitresModel}
	\begin{tabular}{l c c c c c c}
	\toprule
	$\rho',\rho''$ width & \multicolumn{3}{c}{$\pi\pi$} & \multicolumn{3}{c}{$\omega\pi$} 
	\\
	$s_c/\si{GeV^2}$ & \num{3} & \num{4} & \num{5} & \num{3} & \num{4} & \num{5}\\ \midrule
	$(\chi^2/\text{dof})_\text{dif}^{07}$ & \num{2.48} & \num{2.34} & \num{2.34} & \num{2.35} & \num{2.22} & \num{2.21}\\
	$(\chi^2/\text{dof})_\text{dif}^{18}$ & \num{2.77} & \num{2.71} & \num{2.71} & \num{2.77} & \num{2.71} & \num{2.71}\\
	$(\chi^2/\text{dof})_\text{tot}^{07+18}$ & \num{0.86} & \num{0.87} & \num{0.87} & \num{0.84} & \num{0.84} & \num{0.84}\\
	$\bar{\alpha}^*_\Omega[a_2]/\si{GeV^{-2}}$ & \num{0.61 \pm 0.18} & \num{0.50 \pm 0.17} & \num{0.50 \pm 0.17} & \num{0.59 \pm 0.17} & \num{0.49 \pm 0.16} & \num{0.50 \pm 0.16} \\
	$M_{\rho'}/\si{\MeV}$ & \num{1510.8 \pm 6.4} & \num{1510.0 \pm 6.3} & \num{1510.0 \pm 6.3}& \num{1625 \pm 21} & \num{1626 \pm 20} & \num{1627 \pm 21}\\
	$\Gamma_{\rho'}(M_{\rho'}^2)/\si{\MeV}$ & \num{363 \pm 27} & \num{363 \pm 26} & \num{363 \pm 26}& \num{679 \pm 108} & \num{680 \pm 110} & \num{690 \pm 110}\\
	$c_{\rho'}/10^{-1}$ & \num{-2.69 \pm 0.20} & \num{-2.79 \pm 0.20} & \num{-2.78 \pm 0.20}& \num{-2.79 \pm 0.25} & \num{-2.90 \pm 0.25} & \num{-2.90 \pm 0.25}\\
	$c_{\rho''}/10^{-2}$ & \num{0.96 \pm 0.39} & \num{0.93 \pm 0.37} & \num{0.93 \pm 0.37}& \num{1.09 \pm 0.42} & \num{1.09 \pm 0.41} & \num{1.06 \pm 0.40}\\
	$\phi_{\rho''}/\si{\radian}$ & \num{4.53 \pm 0.27} & \num{4.58 \pm 0.25} & \num{4.58 \pm 0.25}& \num{2.52 \pm 0.27} & \num{2.50 \pm 0.25} & \num{2.52 \pm 0.25}\\
	$M_{\rho''}/\si{\MeV}$ & \num{1816.6 \pm 9.9} & \num{1816.5 \pm 9.3} & \num{1816.4 \pm 9.3}& \num{1827 \pm 13} & \num{1825 \pm 12} & \num{1825 \pm 12}\\
	$\Gamma_{\rho''}(M_{\rho''}^2)/\si{\MeV}$ & \num{105 \pm 28} & \num{100 \pm 28} & \num{100 \pm 27}& \num{123 \pm 33} & \num{120 \pm 32} & \num{118 \pm 31}\\
	$N_{07}/10^1\,\si{\text{arb. units}}$ & \num{4.86 \pm 0.18} & \num{4.87 \pm 0.18} & \num{4.88 \pm 0.18}& \num{4.52 \pm 0.18} & \num{4.59 \pm 0.18} & \num{4.59 \pm 0.18}\\
	$N_{18}/10^1\,\si{\text{arb. units}}$ & \num{9.62 \pm 0.21} & \num{9.67 \pm 0.21} & \num{9.67 \pm 0.21}& \num{9.65 \pm 0.21} & \num{9.67 \pm 0.21} & \num{9.67 \pm 0.21}\\
	\bottomrule
	\end{tabular}
\renewcommand{\arraystretch}{1.0}
\end{table*}
\begin{figure*}[htb] 
    \centering
    \begin{minipage}{0.49\textwidth}
        \includegraphics*[width=\textwidth]{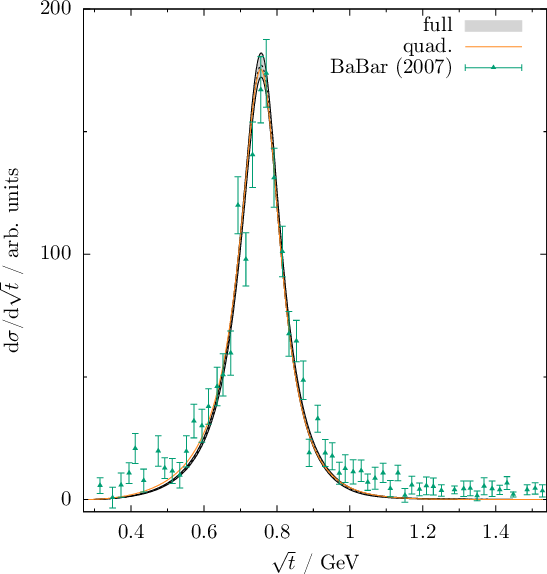}
    \end{minipage}
    \begin{minipage}{0.49\textwidth}
        \includegraphics*[width=\textwidth]{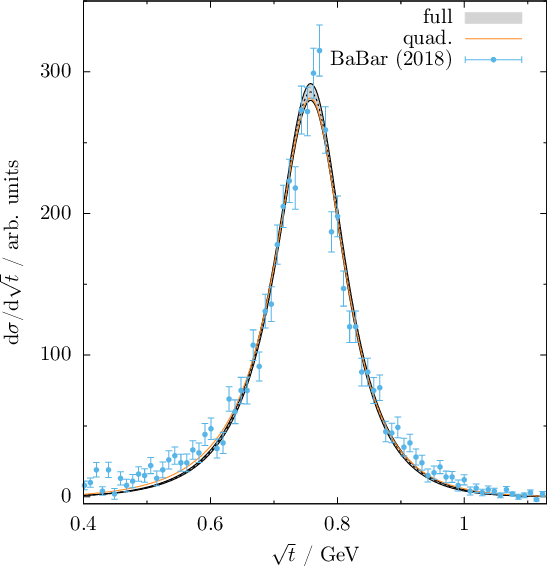}
    \end{minipage}
    \caption{Fit to the differential cross sections from BaBar 2007~\cite{Aubert:2007ef} and 2018~\cite{TheBABAR:2018vvb} up to $\sqrt{t}=\SI{1}{\GeV}$ for $s_c=\SI{3}{\GeV^2}$ (full).  The results shown here correspond to the fits employing the full model, with $\pi\pi$ width parameterizations for the $\rho'$ and $\rho''$ resonances in the total cross section. For comparison, the central value of the fit of the quadratic model (quad.) with $s_p=\SI{1}{\GeV^2}$ and $\pi \pi$ width is also shown.}
    \label{Fig:diffxsec}
\end{figure*}

\begin{figure}[t!] 
    \centering
    \includegraphics[width=\linewidth]{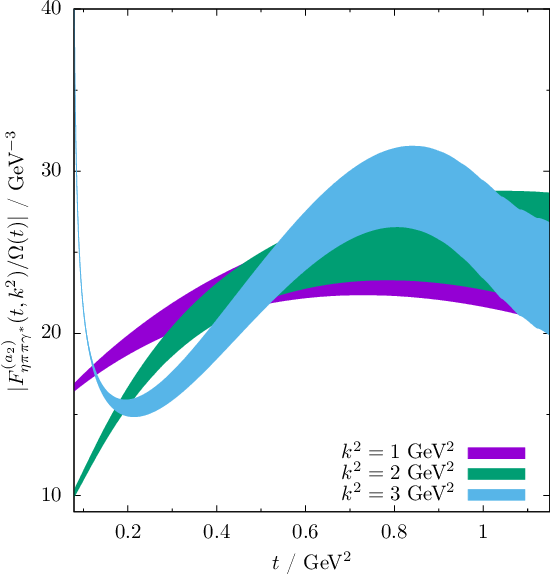}
    \caption{The function $P(t,k^2)$, \textit{cf.}\ Eq.~\eqref{eq:defP}, of the full model for $s_c = \SI{3}{\GeV^2}$ for different $k^2$.}
    \label{Fig:poly_full_ksqvar}
\end{figure}

Here, the full representation of $F_{\eta\pi\pi\gamma^*}(t,k^2)$ in Eq.~\eqref{eq:semifactorization} is used for the $e^+ e^- \to \eta \pi^+ \pi^-$ fits, \textit{i.e.}, including the $k^2$-dependent $a_2$-induced left-hand-cut contribution. The parameter $A=\SI{17.2 \pm 0.4}{\GeV^{-3}}$ in Eq.~\eqref{Eq:Fa2ksq} is fixed by fitting the representation in Eq.~\eqref{eq:f1a2} to $\eta \to \pi^+\pi^- \gamma$ decay data measured by KLOE~\cite{Babusci:2012ft}. The fit results for cutoff parameters $s_c \in \lbrace 3,\, 4,\, 5 \rbrace \si{\GeV^2}$ according to the high-energy prescription in Eq.~\eqref{Eq:fullmodel_highenergy} are summarized in Table~\ref{Tab:FitresModel}. 
We observe that the $\omega\pi$ parameterizations of the $\rho'$ and $\rho''$ widths result in a slightly better fit quality.
The mass and width parameters of the $\rho'$ and $\rho''$, which must not be confused with pole parameters, vary strongly between both descriptions of the energy-dependent width, most prominently for the $\rho'$. 
A comparison with the PDG parameters ($M_{\rho'}^\text{PDG}=\SI{1465 \pm 25}{\MeV}$, $\Gamma_{\rho'}^\text{PDG}=\SI{400 \pm 60}{\MeV}$) is therefore not particularly meaningful.
However, the linear subtraction constant, the central result for our investigation of potential factorization breaking, only varies within the error margins between the two fits.
Additionally, the resulting spectra for cutoff parameter $s_c = 3 \GeV^2$ are shown in Figs.~\ref{Fig:totxsec} and~\ref{Fig:diffxsec}, where the error bands are generated by the fit uncertainties taking their correlations into account.
The description of differential spectra turns out to be slightly worse than for the linear/quadratic models. Those, however, do not capture the nontrivial $k^2$-dependence of the amplitude illustrated in Fig.~\ref{Fig:poly_full_ksqvar}: the change of the left-hand-cut contribution with $k^2$ is seen to be quite significant.
In Fig.~\ref{Fig:diffxsec} the fit result of the full model can be compared to the central value of the quadratic one. Differences between the two curves can only be seen at the lower flank of the $\rho$ resonance. Moreover, if the uncertainty band were added, it would overlap with the curve of the full model almost everywhere.
The large values of the reduced $\chi^2$ of the differential cross section in the full model fit are primarily caused by the flanks of the $\rho$ resonance, as one can see in Fig.~\ref{Fig:diffxsec}. Especially below the $\rho$ peak, the $\chi^2$ receives a  large contribution.  The tension specifically for these $\pi\pi$ invariant masses is surprising, as the dispersive representation should be perfectly reliable there.  Doubly-differential data (in the variables $t$ and $k^2$) would be highly desirable to better understand the origin of this discrepancy.

The Breit--Wigner coupling of the $\rho'$ resonance comes out consistently in all fits; in particular, in contrast to mass and width, it only changes mildly between the two different parameterizations of the energy-dependent width.
Moreover, to good accuracy it fulfills the expectation
\begin{equation}
    \frac{c_\rho'}{c_{\rho}} = - \frac{M_{\rho}^2}{M_{\rho'}^2}  \approx -\num{0.28} 
\end{equation}
(where in our convention Eq.~\eqref{Eq:BWsum}, $c_\rho=1$),
derived from a superconvergence relation for $V \to P$ transition form factors that are expected to drop $\propto k^{-4}$ asymptotically~\cite{Farrar:1975yb,Vainshtein:1977db,Lepage:1979zb,Lepage:1980fj}. The coupling to the $\rho''$ resonance comes out complex, however it is two orders of magnitude suppressed compared to the $\rho$ and more than one order of magnitude with respect to the $\rho'$ coupling. Also the mass and width parameters of the $\rho''$ deviate somewhat from the PDG values, $M_{\rho''}^\text{PDG} = \SI{1720 \pm 20}{\MeV}$ and $\Gamma_{\rho''}^\text{PDG} = \SI{250 \pm 100}{\MeV}$, which may be explained by the occurrence of higher resonances in the spectrum that are not included in our parameterization.

\begin{figure}[t!] 
    \centering
    \includegraphics[width=\linewidth]{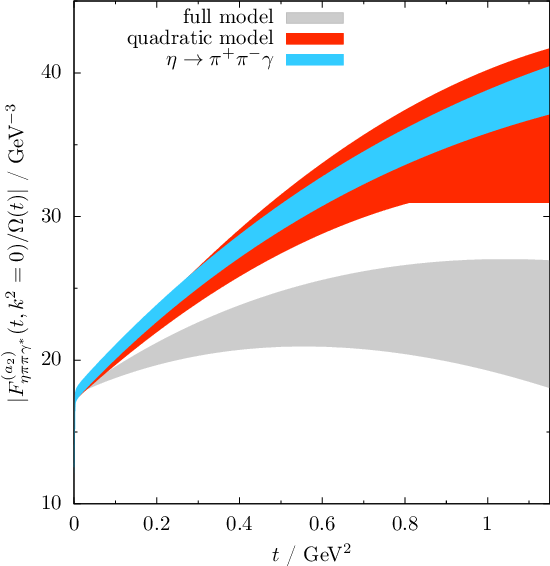}
    \caption{Comparison of the functions $P(t,k^2=0)$, \textit{cf.}\ Eq.~\eqref{eq:defP}, extrapolated from $e^+e^- \to \eta \pi^+\pi^-$ using both the full and the quadratic model, and the one directly obtained from $\eta \to \pi^+\pi^- \gamma$. The band for the full model includes the uncertainty due to $\bar{\alpha}_\Omega^*[a_2]$ as well as the different values for $s_c \in \lbrace 3,\, 4,\, 5 \rbrace\, \si{\GeV^2}$; see Table~\ref{Tab:FitresModel}. In the same way the band for the quadratic model comprises the uncertainty due to $\bar{\alpha}_\Omega^*[a_2]$ as well as $\sqrt{s_p} \in \lbrace 0.9,\, 1.0,\, 1.2 \rbrace\, \si{\GeV}$; see Table~\ref{Tab:FitresPolynomial}. The onset of the high-energy prescription can be seen in the quadratic band around $t = \SI{0.81}{\GeV^2}$.}
    \label{Fig:poly_comp}
\end{figure}
The linear subtraction constant $\bar{\alpha}^*_\Omega[a_2]$ comes out consistently for different cutoff parameters $s_c$, but deviates from the linear constant found in the polynomial part fitted to $\eta \to \pi^+\pi^- \gamma$ decay data of $\SI{1.42\pm0.06}{\GeV^{-2}}$. This is illustrated in Fig.~\ref{Fig:poly_comp}, where the $e^+e^- \to \eta \pi^+\pi^-$ form factor in the extrapolation to $k^2=0$ is compared to the representation in Eq.~\eqref{eq:f1a2} fitted to $\eta \to \pi^+\pi^- \gamma$ decay data. This is in contrast to the quadratic model, which allows for a smooth connection between the extrapolated $e^+e^- \to \eta \pi^+\pi^-$ and the $\eta$ decay data.  We therefore have to conclude that the dominant left-hand-cut contribution of the $a_2$ exchange provides a factorization breaking mechanism that is too strong compared to the available data, and likely calls for further effects that mitigate this to some extent.
\esp

\section{Consequences for the $\eta$ transition form factor}\label{sec:etaTFF}

\begin{figure} 
    \centering
    \includegraphics[width=\linewidth]{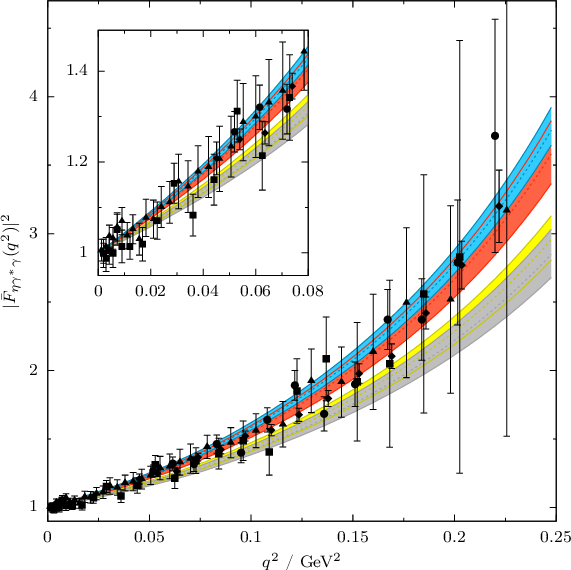}
    \caption{Results for the normalized singly-virtual $\eta$ transition form factor. The very small isoscalar contribution has been omitted. We show $\eta \to e^+ e^- \gamma$ data from the A2 (squares~\cite{Aguar-Bartolome:2013vpw}, triangles~\cite{Adlarson:2016hpp}) and $\eta \to \mu^+ \mu^- \gamma$ data from the NA60 (circles~\cite{Arnaldi:2009aa}, diamonds~\cite{Arnaldi:2016pzu}) collaborations. The insert magnifies the low-$q^2$ region. The gray band shows the result from the full model for $s_c = \SI{3}{GeV^2}$. The red (yellow) band represents the outcome of the quadratic (linear) model for $s_p = \SI{1}{\GeV^2}$. Additionally, for comparison, the result derived from $\eta \to \pi^+ \pi^- \gamma$ is shown as the blue band.  The central results are represented by dashed lines colored accordingly throughout.}
    \label{Fig:singly_tff}
\end{figure}
\bsp
The observations made on the different models for $F_{\eta\pi\pi\gamma^*}(t,k^2)$ in the previous section immediately translate into the $\eta$ transition form factor. 
Figure~\ref{Fig:singly_tff} shows the results for the normalized $\eta \to \gamma \gamma^*$ transition form factor based on a once-subtracted variant of Eq.~\eqref{eq:DR} in the time-like regime (thus enforcing the correct normalization by hand), compared to experimental data~\cite{Aguar-Bartolome:2013vpw,Adlarson:2016hpp,Arnaldi:2009aa,Arnaldi:2016pzu}.
Here, the dispersion integral has been truncated~\cite{Hanhart:2013vba}, with the upper limit of integration varied between $M_{\eta'}^2$ and \SI{2}{\GeV^2}. The error bands in Fig.~\ref{Fig:singly_tff} are generated from the error of the linear subtraction constant $\bar{\alpha}_\Omega^*[a_2]$, the variation of the integration upper limit, the pion vector form factor, and the errors of the branching ratios entering the result. The central results shown as the dashed lines stem from evaluating the integral with cutoff \SI{1}{\GeV^2}.  Not surprisingly, the transition form factor based on the extrapolated quadratic model agrees well with the direct calculation based on $\eta\to\pi^+\pi^-\gamma$, while the extrapolated linear and full models yield a form factor significantly below the others. This is also illustrated by the $\chi^2/\text{dof}$, where $\chi^2$ as usual refers to the weighted sum of residuals between data and the central values of the predictions and $\text{dof}$ stands for the number of data points (as there are no fit parameters to be adjusted). 
The full model yields $\chi^2_\text{full}/\text{dof} = 2.09$, when compared to the linear model with $\chi^2_\text{lin.}/\text{dof} = 1.54$ and the quadratic model with $\chi^2_\text{quad.}/\text{dof} = 0.48$.  This is to be compared to $\chi^2_{\pi\pi\gamma}/\text{dof} = 0.78$ for the TFF calculated directly based on $\eta\to\pi^+\pi^-\gamma$ decay data.
In particular the data from Refs.~\cite{Arnaldi:2009aa,Arnaldi:2016pzu} disfavor the TFF based on the extrapolated full model rather strongly.

These differences are also reflected in the slope parameter of the TFF defined by
\begin{equation}
    \bar{F}_{\eta\gamma^*\gamma}(q^2) = 1 + b_\eta q^2 + {\cal O}(q^4) \,.
\end{equation}
According to Eq.~\eqref{eq:DR}, its isovector part fulfills a sum rule 
\begin{equation}
b_\eta = \frac{1}{96 \pi^{2} F_{\eta\gamma\gamma}} \int_{4 M_\pi^2} ^\infty \frac{\diff t}{t} 
         \sigma_\pi^3(t) F_{\eta\pi\pi\gamma^*}(t,0) [F_\pi^V(t)]^* \,.
    \label{eq:b_eta}
\end{equation}
The slope parameters of the respective models are listed in Table~\ref{Tab:slope_par}.
\begin{table} 
\renewcommand{\arraystretch}{1.5}
	\centering
	\caption{Slope parameter results for the different models, where the small isoscalar contribution has been omitted; see Eq.~\eqref{eq:b_eta}.}
    \begin{tabular}{c c}
        \toprule
        model & $b_\eta/\si{\GeV^{-2}}$\\ \midrule
        linear &  $1.6^{+0.1}_{-0.1}$\\
        quadratic & $1.9^{+0.2}_{-0.1}$\\
        full & $1.5^{+0.1}_{-0.1}$\\ \midrule
        $\eta \to \pi^+ \pi^- \gamma$~\cite{Kubis:2015sga} & $1.9^{+0.2}_{-0.1}$\\ \bottomrule
    \end{tabular}
    \label{Tab:slope_par}
\end{table}
The isoscalar contribution $b_\eta^{(I=0)}=-\SI{0.022}{\GeV^{-2}}$~\cite{Hanhart:2013vba} is inside the quoted uncertainties throughout.
\esp
\begin{figure} 
    \centering
    \includegraphics[width=\linewidth]{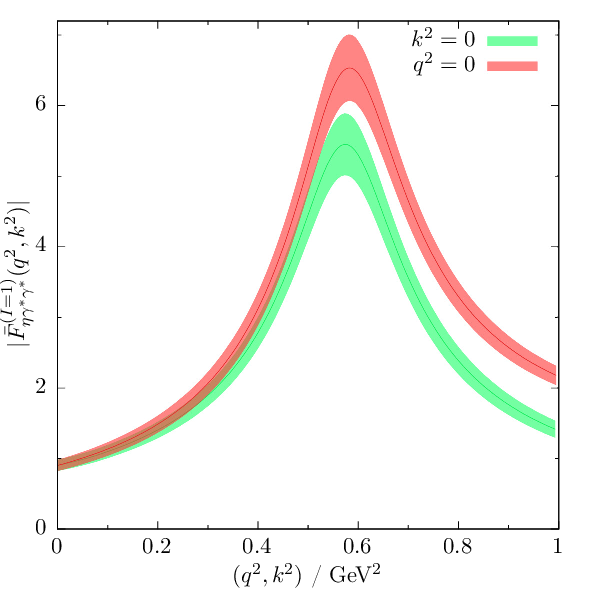}
    \caption{Comparison of the normalized isovector $\eta$ transition form factor $\bar F_{\eta\gamma^*\gamma^*}^{(I=1)}(q^2,k^2)$ in the singly-virtual limits $k^2=0$ and $q^2=0$. Results are shown from the full model for $s_c = \SI{3}{\GeV^2}$. The error band for the $k^2=0$ case is generated as described before, whereas the error band of the $q^2=0$ curve additionally takes the fit errors of the parameters in $\tilde{F}_{\eta \gamma \gamma^*}(k^2)$ and their correlations into account.}
    \label{Fig:TFF_qk_symm}
\end{figure}
Bose symmetry requires the TFF as well as its isovector part to be symmetric under the exchange of the two arguments $q^2$ and $k^2$.  This symmetry is by no means manifest in our model, which in contrast is built onto information in two disjoint kinematic ranges: the dispersion relation~\eqref{eq:DR} is designed for the region $q^2\leq1\GeV^2$, while the dependence on the second variable is fitted in the range $1\GeV^2 \leq k^2 \leq (1.9\GeV)^2$.  Indeed, Fig.~\ref{Fig:TFF_qk_symm} demonstrates that the two different singly-virtual limits $\bar F_{\eta\gamma^*\gamma^*}^{(I=1)}(q^2,k^2=0)$ and $\bar F_{\eta\gamma^*\gamma^*}^{(I=1)}(q^2=0,k^2)$ do not coincide in the region of the $\rho(770)$ peak, although the disagreement hardly exceeds the combined width of the two uncertainty bands.  
Obviously, in order to calculate and predict the limit $q^2=0$, we are required to use the unsubtracted dispersion relation~\eqref{eq:DR}, without the integral being cut off. Instead, the high-energy prescription in Eq.~\eqref{Eq:fullmodel_highenergy} has been used above a certain cutoff varied between $1$ and $\SI{2}{\GeV^2}$. Note that the TFF shown in Fig.~\ref{Fig:TFF_qk_symm} is normalized using $F_{\eta\gamma\gamma}$ based on the physical value of the two-photon partial width.   
$\bar F_{\eta\gamma^*\gamma^*}^{(I=1)}(0,0)$ is hence a test of the hypothesis that two-pion intermediate states saturate the isovector dispersion relation to a large extent: indeed, the normalization is reproduced to $90(8)\%$.  Figure~\ref{Fig:TFF_qk_symm} suggests that the remaining deficit, which is due to omitted heavier intermediate states (such as $4\pi$, $\pi\omega$, $K\bar K$, etc.), likely increases somewhat around the $\rho(770)$, but not dramatically so.

\section{Summary and discussion}
\bsp
In this article, we have investigated the consistency of pion--pion spectra in the closely related reactions $\eta\to\pi^+\pi^-\gamma$ and $e^+e^-\to\eta\pi^+\pi^-$, which serve as central input to dispersion-theoretical analyses of the singly- and doubly-virtual $\eta$ transition form factor, respectively.  This comparison has been quantified in terms of the residual dependence on the $\pi\pi$ invariant mass squared $t$ that multiplies universal elastic rescattering, as encapsulated in the Omnès function.  While the small phase space available in the decay $\eta\to\pi^+\pi^-\gamma$ is sufficiently accurately described by a residual linear $t$-dependence~\cite{Stollenwerk:2011zz,Babusci:2012ft}, both a study of the dominant left-hand cuts, induced by crossed-channel exchanges of the $a_2$ tensor meson~\cite{Kubis:2015sga}, and data on the closely related $\eta'\to\pi^+\pi^-\gamma$ decay~\cite{Hanhart:2016pcd,Ablikim:2017fll} demonstrate that at least a quadratic polynomial is required to describe the $\pi\pi$ spectra up to $1\GeV^2$.  

The present analysis of the pion--pion distributions in $e^+e^-\to\eta\pi^+\pi^-$ confirms that, indeed, a linear parameterization is insufficient to describe these with a universal slope parameter independently of the dilepton squared invariant mass $k^2$.  On the other hand, a quadratic polynomial, with the quadratic term in $t$ fixed from the size of curvature induced by the left-hand cut at $k^2=0$, seems well compatible with a na\"ive factorization assumption for the $t$ and $k^2$ dependences, and as a result, for the doubly-virtual $\eta$ transition form factor varying with its two arguments according to Eq.~\eqref{eq:factor}.  The somewhat puzzling observation is that the natural breaking of factorization, induced by the nontrivial $k^2$ dependence of the $a_2$ exchange contribution, is too strong to be consistent with the data: the amplitude we have built on this physically well-motivated picture does not link the available $e^+e^-\to\eta\pi^+\pi^-$ data to those for the $\eta$ decay in a consistent manner; factorization breaking has to be mitigated significantly by other mechanisms not investigated so far.

One way how such a damping might be achieved is through different $k^2$-dependences of the two considered isobar components in the $\gamma^*\to\eta\pi^+\pi^-$ amplitude: we have implicitly assumed that the dominant ``$\eta\rho$'' and the supplementary $a_2\pi$ contribution responsible for the left-hand cut evolve with $k^2$ in parallel.  The only way to understand deviations from this simple picture is to analyze differential distributions in $\eta\pi^+\pi^-$ in different bins of $k^2$, as opposed to the $k^2$-integrated $\pi\pi$ spectra available to date~\cite{Aubert:2007ef,TheBABAR:2018vvb,Gribanov:2019qgw}.  Such information, combined with a doubly-dispersive construction based on a $\eta\to2(\pi^+\pi^-)$ amplitude at low energies (\textit{cf.}\ the corresponding $\eta'\to2(\pi^+\pi^-)$ decays~\cite{Guo:2011ir,Ablikim:2014eoc} and the discussion in Ref.~\cite{Gan:2020aco}), would then pave the way towards a more complete understanding of the doubly-virtual $\eta$ transition form factor.
\esp

\begin{acknowledgements}
\bsp
We are grateful to the anonymous referee who pointed out a conceptual inconsistency in the first version of this manuscript.
We would like to thank Eugeny P.\ Solodov and Simon Eidelman 
for helpful discussions and advice, and Evgeny A.\ Kozyrev for providing us with the data of Ref.~\cite{TheBABAR:2018vvb}.  
Furthermore, we are grateful to Peter Stoffer for help with the pion--pion phase shift solution of Ref.~\cite{Caprini:2011ky}.
This work is supported in part by the DFG and the NSFC through funds provided to the Sino--German Collaborative Research Center TRR110 ``Symmetries and the Emergence of Structure in QCD'' (DFG Project-ID 196253076, NSFC Grant No.~12070131001). The project that gave rise to these results received the support of a fellowship from ``la Caixa'' Foundation (ID 100010434), fellowship code LCF/BQ/IN17/11620037.
\esp
\end{acknowledgements}

\appendix

\section{Technical implementation}

\subsection{$P$-wave phase shift}\label{app:Pwave}

\begin{figure} 
    \centering
    \includegraphics[width=\linewidth]{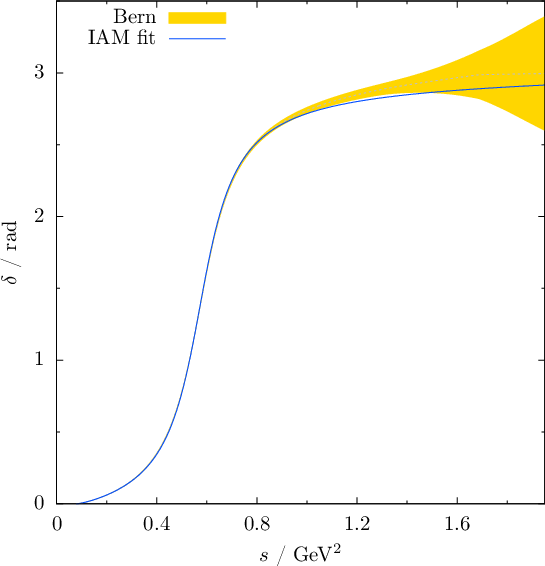}
    \caption{Fit of the IAM phase to the phase of Ref.~\cite{Caprini:2011ky} (``Bern analysis'', with the dashed curve as the central result) up to $\sqrt{s} = \SI{1.3}{\GeV}$.}
    \label{Fig:PhaseFit}
\end{figure}

As mentioned in the main text, the $\pi \pi$ phase shift employed in this work is calculated via the inverse-amplitude method (IAM) from an approximation of the two-loop representation of Ref.~\cite{Niehus:2020gmf}. The amplitude for $\pi \pi \to \pi \pi$ scattering in chiral perturbation theory (ChPT) up to next-to-leading order (NLO), projected onto the $P$-wave, can be written as
\begin{equation}
    t_\text{ChPT}(s) = t_2(s) + t_4(s) + \mathcal{O}(p^6) \,,
\end{equation}
where $t_i$ labels the contribution of chiral order $p^i$. The $P$-wave projected amplitudes up to NLO are given by~\cite{Gasser:1983yg,Dax:2018rvs}
\begin{align}
        t_2(s)&=\frac{s\sigma^2}{96\pi F^2}\,, \notag\\
        t_4(s)&=\frac{t_{2}(s)}{48\pi^2F^2}\bigg\lbrace s \left(\bar{l}_2 - \bar{l}_1+\frac{1}{3}\right)-\frac{15}{2}M_\pi^2 \notag\\
        &\qquad -\frac{M_\pi^4}{2s}\Big[41-2L_\sigma\big(73-25\sigma^2\big) \notag\\
        &\qquad+3L_\sigma^2\big(5-32\sigma^2+3\sigma^4\big) \Big]\bigg\rbrace+i\sigma\,(t_2(s))^2\,,
\end{align}
where 
\begin{equation}
    L_\sigma=\frac{1}{\sigma^2}\left(\frac{1}{2\sigma}\log\frac{1+\sigma}{1-\sigma}-1\right)
\end{equation}
and $\sigma = \sigma_\pi(s)$. In those expressions $F$ represents the pion decay constant in the chiral limit, which can be extracted from the ratio $F_\pi/F = \num{1.062 \pm 0.007}$~\cite{Aoki:2019cca,Bazavov:2010hj,Borsanyi:2012zv,Durr:2013goa,Boyle:2015exm,Beane:2011zm}. The combination of low-energy constants $\bar{l}_2 - \bar{l}_1$ is later used as a free parameter. Utilizing the IAM, we write the unitarized amplitude as~\cite{Dobado:1989qm,Truong:1991gv,Dobado:1992ha}
\begin{equation}
    t_\text{IAM}(s) = \frac{(t_2(s))^2}{t_2(s) - t_4(s)} \,.
\end{equation}
Now the $\pi \pi$ $P$-wave phase shift can be extracted from this expression as $\delta_1^1 = \operatorname{arg}\, t_\text{IAM}(s)$. However, in order to enforce the phase to converge to $\pi$ for $s \to \infty$, we add terms, inspired by the next-to-next-to leading order amplitude, to $t_4$ by hand:
\begin{equation}
    t_4(s) \mapsto t_4(s) + \frac{t_{2}(s)}{48\pi^2F^2}\big(\hat{l}_s s^2 + \hat{l}_\pi M_\pi^4\big) \,,
\end{equation}
where two additional parameters $\hat{l}_s$ and $\hat{l}_\pi$ have been introduced. In order to fix the three parameters, we fit the phase parameterization from the IAM amplitude to the solution of the Roy-equation analysis of $\pi \pi$ scattering~\cite{Caprini:2011ky}. The fit up to $\sqrt{s} = \SI{1.3}{\GeV}$ is displayed in Fig.~\ref{Fig:PhaseFit}. The parameters resulting from this fit are $\bar{l}_2-\bar{l}_1 = \num{4.47 \pm 0.03}$, $\hat{l}_s = \SI{1.74 \pm 0.03}{\GeV^{-2}}$, and $\hat{l}_\pi = \SI{560 \pm 20}{\GeV^{-2}}$.

\subsection{Hat function}\label{app:hat}
The analytical hat function obtained through the angular integral over the $a_2$ exchange amplitudes in Eq.~ \eqref{Eq:hat_function} is implemented in a smooth way by choosing the correct sheets of the logarithm appearing in $Q(y)$, explicitly given by
\begin{align}
    &Q(y) = \\
    &y \bigg[ \frac{1-y^2}{2} \left( \log \left|\frac{y+1}{y-1}\right| + i ~\text{arg} \left( \frac{y+1}{y-1}\right) + i~ 2\pi \theta \right) +y \bigg] \,, \nonumber 
\end{align}
where
\begin{equation}
    \theta =
    \begin{cases}
        1 & \text{for} \quad \text{Im} \left( \frac{y+1}{y-1} \right) < 0 \quad \text{and} \\
        &  \qquad \Bigg( k^2>\frac{\left(M_{a_2}^2-M_\pi^2\right)^2}{M_\eta^2} ~ \text{or}~\text{Re} \left( \frac{y+1}{y-1} \right) < 0  \Bigg) \,,\\
        -1 & \text{for} \quad \text{Im} \left( \frac{y+1}{y-1} \right)> 0 \quad \text{and}  \\
        &\qquad \Bigg( k^2>\frac{\left(M_{a_2}^2-M_\pi^2\right)^2}{M_\eta^2} ~ \text{or}~\text{Re} \left( \frac{y+1}{y-1} \right) < 0  \Bigg) \,,\\
        0 & \text{else}\,.
    \end{cases}
\end{equation}

\bibliographystyle{utphysmod}
\bibliography{Literature}

\providecommand{\href}[2]{#2}\begingroup\raggedright\begin{thebibliography}{10}

\bibitem{Colangelo:2014dfa}
G.~Colangelo, M.~Hoferichter, M.~Procura, and P.~Stoffer,
  \href{http://dx.doi.org/10.1007/JHEP09(2014)091}{JHEP {\bfseries 09}, 091
  (2014)} [\href{https://arxiv.org/abs/1402.7081}{{arXiv:1402.7081 [hep-ph]}}].

\bibitem{Colangelo:2014pva}
G.~Colangelo, M.~Hoferichter, B.~Kubis, M.~Procura, and P.~Stoffer,
  \href{http://dx.doi.org/10.1016/j.physletb.2014.09.021}{Phys. Lett. B
  {\bfseries 738}, 6 (2014)}
  [\href{https://arxiv.org/abs/1408.2517}{{arXiv:1408.2517 [hep-ph]}}].

\bibitem{Colangelo:2015ama}
G.~Colangelo, M.~Hoferichter, M.~Procura, and P.~Stoffer,
  \href{http://dx.doi.org/10.1007/JHEP09(2015)074}{JHEP {\bfseries 09}, 074
  (2015)} [\href{https://arxiv.org/abs/1506.01386}{{arXiv:1506.01386
  [hep-ph]}}].

\bibitem{Aoyama:2020ynm}
T.~Aoyama {\em et~al.},
  \href{http://dx.doi.org/10.1016/j.physrep.2020.07.006}{Phys. Rept. {\bfseries
  887}, 1 (2020)} [\href{https://arxiv.org/abs/2006.04822}{{arXiv:2006.04822
  [hep-ph]}}].

\bibitem{Pauk:2014rfa}
V.~Pauk and M.~Vanderhaeghen,
  \href{http://dx.doi.org/10.1103/PhysRevD.90.113012}{Phys. Rev. D {\bfseries
  90}, 113012 (2014)} [\href{https://arxiv.org/abs/1409.0819}{{arXiv:1409.0819
  [hep-ph]}}].

\bibitem{Bennett:2006fi}
G.~W. Bennett {\em et~al.} [Muon $g-2$ Collaboration],
  \href{http://dx.doi.org/10.1103/PhysRevD.73.072003}{Phys. Rev. D {\bfseries
  73}, 072003 (2006)}
  [\href{https://arxiv.org/abs/hep-ex/0602035}{{arXiv:hep-ex/0602035}}].

\bibitem{Abi:2021gix}
B.~Abi {\em et~al.} [Muon $g-2$ Collaboration],
  \href{http://dx.doi.org/10.1103/PhysRevLett.126.141801}{Phys. Rev. Lett.
  {\bfseries 126}, 141801 (2021)}
  [\href{https://arxiv.org/abs/2104.03281}{{arXiv:2104.03281 [hep-ex]}}].

\bibitem{Albahri:2021ixb}
T.~Albahri {\em et~al.} [Muon $g-2$ Collaboration],
  \href{http://dx.doi.org/10.1103/PhysRevD.103.072002}{Phys. Rev. D {\bfseries
  103}, 072002 (2021)}
  [\href{https://arxiv.org/abs/2104.03247}{{arXiv:2104.03247 [hep-ex]}}].

\bibitem{Aoyama:2012wk}
T.~Aoyama, M.~Hayakawa, T.~Kinoshita, and M.~Nio,
  \href{http://dx.doi.org/10.1103/PhysRevLett.109.111808}{Phys. Rev. Lett.
  {\bfseries 109}, 111808 (2012)}
[\href{https://arxiv.org/abs/1205.5370}{{arXiv:1205.5370 [hep-ph]}}].

\bibitem{Aoyama:2019ryr}
T.~Aoyama, T.~Kinoshita, and M.~Nio,
  \href{http://dx.doi.org/10.3390/atoms7010028}{Atoms {\bfseries 7}, 28
  (2019)}.

\bibitem{Czarnecki:2002nt}
A.~Czarnecki, W.~J. Marciano, and A.~Vainshtein,
  \href{http://dx.doi.org/10.1103/PhysRevD.67.073006}{Phys. Rev. {\bfseries
  D67}, 073006 (2003)}
  [\href{https://arxiv.org/abs/hep-ph/0212229}{{arXiv:hep-ph/0212229}}],
[Erratum: Phys. Rev. {\bf D73}, 119901 (2006)].

\bibitem{Gnendiger:2013pva}
C.~Gnendiger, D.~St{\"o}ckinger, and H.~St{\"o}ckinger-Kim,
  \href{http://dx.doi.org/10.1103/PhysRevD.88.053005}{Phys. Rev. {\bfseries
  D88}, 053005 (2013)}
[\href{https://arxiv.org/abs/1306.5546}{{arXiv:1306.5546 [hep-ph]}}].

\bibitem{Davier:2017zfy}
M.~Davier, A.~Hoecker, B.~Malaescu, and Z.~Zhang,
  \href{http://dx.doi.org/10.1140/epjc/s10052-017-5161-6}{Eur. Phys. J.
  {\bfseries C77}, 827 (2017)}
[\href{https://arxiv.org/abs/1706.09436}{{arXiv:1706.09436 [hep-ph]}}].

\bibitem{Keshavarzi:2018mgv}
A.~Keshavarzi, D.~Nomura, and T.~Teubner,
  \href{http://dx.doi.org/10.1103/PhysRevD.97.114025}{Phys. Rev. {\bfseries
  D97}, 114025 (2018)}
[\href{https://arxiv.org/abs/1802.02995}{{arXiv:1802.02995 [hep-ph]}}].

\bibitem{Colangelo:2018mtw}
G.~Colangelo, M.~Hoferichter, and P.~Stoffer,
  \href{http://dx.doi.org/10.1007/JHEP02(2019)006}{JHEP {\bfseries 02}, 006
  (2019)}
[\href{https://arxiv.org/abs/1810.00007}{{arXiv:1810.00007 [hep-ph]}}].

\bibitem{Hoferichter:2019gzf}
M.~Hoferichter, B.-L. Hoid, and B.~Kubis,
  \href{http://dx.doi.org/10.1007/JHEP08(2019)137}{JHEP {\bfseries 08}, 137
  (2019)}
[\href{https://arxiv.org/abs/1907.01556}{{arXiv:1907.01556 [hep-ph]}}].

\bibitem{Davier:2019can}
M.~Davier, A.~Hoecker, B.~Malaescu, and Z.~Zhang,
  \href{http://dx.doi.org/10.1140/epjc/s10052-020-7792-2}{Eur. Phys. J.
  {\bfseries C80}, 241 (2020)}
  [\href{https://arxiv.org/abs/1908.00921}{{arXiv:1908.00921 [hep-ph]}}],
[Erratum: Eur. Phys. J. {\bf C80}, 410 (2020)].

\bibitem{Keshavarzi:2019abf}
A.~Keshavarzi, D.~Nomura, and T.~Teubner,
  \href{http://dx.doi.org/10.1103/PhysRevD.101.014029}{Phys. Rev. {\bfseries
  D101}, 014029 (2020)}
[\href{https://arxiv.org/abs/1911.00367}{{arXiv:1911.00367 [hep-ph]}}].

\bibitem{Hoid:2020xjs}
B.-L. Hoid, M.~Hoferichter, and B.~Kubis,
  \href{http://dx.doi.org/10.1140/epjc/s10052-020-08550-2}{Eur. Phys. J. C
  {\bfseries 80}, 988 (2020)}
  [\href{https://arxiv.org/abs/2007.12696}{{arXiv:2007.12696 [hep-ph]}}].

\bibitem{Kurz:2014wya}
A.~Kurz, T.~Liu, P.~Marquard, and M.~Steinhauser,
  \href{http://dx.doi.org/10.1016/j.physletb.2014.05.043}{Phys. Lett.
  {\bfseries B734}, 144 (2014)}
[\href{https://arxiv.org/abs/1403.6400}{{arXiv:1403.6400 [hep-ph]}}].

\bibitem{Melnikov:2003xd}
K.~Melnikov and A.~Vainshtein,
  \href{http://dx.doi.org/10.1103/PhysRevD.70.113006}{Phys. Rev. {\bfseries
  D70}, 113006 (2004)}
[\href{https://arxiv.org/abs/hep-ph/0312226}{{arXiv:hep-ph/0312226}}].

\bibitem{Masjuan:2017tvw}
P.~Masjuan and P.~S{\'a}nchez-Puertas,
  \href{http://dx.doi.org/10.1103/PhysRevD.95.054026}{Phys. Rev. {\bfseries
  D95}, 054026 (2017)}
[\href{https://arxiv.org/abs/1701.05829}{{arXiv:1701.05829 [hep-ph]}}].

\bibitem{Colangelo:2017qdm}
G.~Colangelo, M.~Hoferichter, M.~Procura, and P.~Stoffer,
  \href{http://dx.doi.org/10.1103/PhysRevLett.118.232001}{Phys. Rev. Lett.
  {\bfseries 118}, 232001 (2017)}
  [\href{https://arxiv.org/abs/1701.06554}{{arXiv:1701.06554 [hep-ph]}}].

\bibitem{Colangelo:2017fiz}
G.~Colangelo, M.~Hoferichter, M.~Procura, and P.~Stoffer,
  \href{http://dx.doi.org/10.1007/JHEP04(2017)161}{JHEP {\bfseries 04}, 161
  (2017)}
[\href{https://arxiv.org/abs/1702.07347}{{arXiv:1702.07347 [hep-ph]}}].

\bibitem{Danilkin:2021icn}
I.~Danilkin, M.~Hoferichter, and P.~Stoffer,
  \href{http://dx.doi.org/10.1016/j.physletb.2021.136502}{Phys. Lett. B
  {\bfseries 820}, 136502 (2021)}
  [\href{https://arxiv.org/abs/2105.01666}{{arXiv:2105.01666 [hep-ph]}}].

\bibitem{Hoferichter:2018dmo}
M.~Hoferichter, B.-L. Hoid, B.~Kubis, S.~Leupold, and S.~P. Schneider,
  \href{http://dx.doi.org/10.1103/PhysRevLett.121.112002}{Phys. Rev. Lett.
  {\bfseries 121}, 112002 (2018)}
  [\href{https://arxiv.org/abs/1805.01471}{{arXiv:1805.01471 [hep-ph]}}].

\bibitem{Hoferichter:2018kwz}
M.~Hoferichter, B.-L. Hoid, B.~Kubis, S.~Leupold, and S.~P. Schneider,
  \href{http://dx.doi.org/10.1007/JHEP10(2018)141}{JHEP {\bfseries 10}, 141
  (2018)} [\href{https://arxiv.org/abs/1808.04823}{{arXiv:1808.04823
  [hep-ph]}}].

\bibitem{Gerardin:2019vio}
A.~G{\'e}rardin, H.~B. Meyer, and A.~Nyffeler,
  \href{http://dx.doi.org/10.1103/PhysRevD.100.034520}{Phys. Rev. {\bfseries
  D100}, 034520 (2019)}
[\href{https://arxiv.org/abs/1903.09471}{{arXiv:1903.09471 [hep-lat]}}].

\bibitem{Bijnens:2019ghy}
J.~Bijnens, N.~Hermansson-Truedsson, and A.~Rodr{\'i}guez-S{\'a}nchez,
  \href{http://dx.doi.org/10.1016/j.physletb.2019.134994}{Phys. Lett.
  {\bfseries B798}, 134994 (2019)}
[\href{https://arxiv.org/abs/1908.03331}{{arXiv:1908.03331 [hep-ph]}}].

\bibitem{Colangelo:2019lpu}
G.~Colangelo, F.~Hagelstein, M.~Hoferichter, L.~Laub, and P.~Stoffer,
  \href{http://dx.doi.org/10.1103/PhysRevD.101.051501}{Phys. Rev. D {\bfseries
  101}, 051501 (2020)}
  [\href{https://arxiv.org/abs/1910.11881}{{arXiv:1910.11881 [hep-ph]}}].

\bibitem{Colangelo:2019uex}
G.~Colangelo, F.~Hagelstein, M.~Hoferichter, L.~Laub, and P.~Stoffer,
  \href{http://dx.doi.org/10.1007/JHEP03(2020)101}{JHEP {\bfseries 03}, 101
  (2020)}
[\href{https://arxiv.org/abs/1910.13432}{{arXiv:1910.13432 [hep-ph]}}].

\bibitem{Blum:2019ugy}
T.~Blum, N.~Christ, M.~Hayakawa, T.~Izubuchi, L.~Jin, C.~Jung, and C.~Lehner,
  \href{http://dx.doi.org/10.1103/PhysRevLett.124.132002}{Phys. Rev. Lett.
  {\bfseries 124}, 132002 (2020)}
[\href{https://arxiv.org/abs/1911.08123}{{arXiv:1911.08123 [hep-lat]}}].

\bibitem{Colangelo:2014qya}
G.~Colangelo, M.~Hoferichter, A.~Nyffeler, M.~Passera, and P.~Stoffer,
  \href{http://dx.doi.org/10.1016/j.physletb.2014.06.012}{Phys. Lett.
  {\bfseries B735}, 90 (2014)}
[\href{https://arxiv.org/abs/1403.7512}{{arXiv:1403.7512 [hep-ph]}}].

\bibitem{Grange:2015fou}
J.~Grange {\em et~al.}, [Muon $g-2$ Collaboration],
  \href{https://arxiv.org/abs/1501.06858}{{arXiv:1501.06858
  [physics.ins-det]}}.

\bibitem{Hoferichter:2020lap}
M.~Hoferichter and P.~Stoffer,
  \href{http://dx.doi.org/10.1007/JHEP05(2020)159}{JHEP {\bfseries 05}, 159
  (2020)} [\href{https://arxiv.org/abs/2004.06127}{{arXiv:2004.06127
  [hep-ph]}}].

\bibitem{Zanke:2021wiq}
M.~Zanke, M.~Hoferichter, and B.~Kubis,
  \href{http://dx.doi.org/10.1007/JHEP07(2021)106}{JHEP {\bfseries 07}, 106
  (2021)} [\href{https://arxiv.org/abs/2103.09829}{{arXiv:2103.09829
  [hep-ph]}}].

\bibitem{Knecht:2020xyr}
M.~Knecht, \href{http://dx.doi.org/10.1007/JHEP08(2020)056}{JHEP {\bfseries
  08}, 056 (2020)} [\href{https://arxiv.org/abs/2005.09929}{{arXiv:2005.09929
  [hep-ph]}}].

\bibitem{Ludtke:2020moa}
J.~L\"udtke and M.~Procura,
  \href{http://dx.doi.org/10.1140/epjc/s10052-020-08611-6}{Eur. Phys. J. C
  {\bfseries 80}, 1108 (2020)}
  [\href{https://arxiv.org/abs/2006.00007}{{arXiv:2006.00007 [hep-ph]}}].

\bibitem{Bijnens:2020xnl}
J.~Bijnens, N.~Hermansson-Truedsson, L.~Laub, and A.~Rodr\'\i{}guez-S\'anchez,
  \href{http://dx.doi.org/10.1007/JHEP10(2020)203}{JHEP {\bfseries 10}, 203
  (2020)} [\href{https://arxiv.org/abs/2008.13487}{{arXiv:2008.13487
  [hep-ph]}}].

\bibitem{Bijnens:2021jqo}
J.~Bijnens, N.~Hermansson-Truedsson, L.~Laub, and A.~Rodr\'\i{}guez-S\'anchez,
  \href{http://dx.doi.org/10.1007/JHEP04(2021)240}{JHEP {\bfseries 04}, 240
  (2021)} [\href{https://arxiv.org/abs/2101.09169}{{arXiv:2101.09169
  [hep-ph]}}].

\bibitem{Gan:2020aco}
L.~Gan, B.~Kubis, E.~Passemar, and S.~Tulin,
  \href{https://arxiv.org/abs/2007.00664}{{arXiv:2007.00664 [hep-ph]}}.

\bibitem{Hanhart:2013vba}
C.~Hanhart, A.~Kup\'s\'c, U.-G. Mei\ss{}ner, F.~Stollenwerk, and A.~Wirzba,
  \href{http://dx.doi.org/10.1140/epjc/s10052-013-2668-3}{Eur. Phys. J. C
  {\bfseries 73}, 2668 (2013)}
  [\href{https://arxiv.org/abs/1307.5654}{{arXiv:1307.5654 [hep-ph]}}],
  [Erratum: Eur. Phys. J. C \textbf{75}, 242 (2015)].

\bibitem{Stollenwerk:2011zz}
F.~Stollenwerk, C.~Hanhart, A.~Kup\'s\'c, U.-G. Mei{\ss}ner, and A.~Wirzba,
  \href{http://dx.doi.org/10.1016/j.physletb.2011.12.008}{Phys. Lett. B
  {\bfseries 707}, 184 (2012)}
  [\href{https://arxiv.org/abs/1108.2419}{{arXiv:1108.2419 [nucl-th]}}].

\bibitem{Adlarson:2011xb}
P.~Adlarson {\em et~al.} [WASA-at-COSY Collaboration],
  \href{http://dx.doi.org/10.1016/j.physletb.2011.12.027}{Phys. Lett. B
  {\bfseries 707}, 243 (2012)}
  [\href{https://arxiv.org/abs/1107.5277}{{arXiv:1107.5277 [nucl-ex]}}].

\bibitem{Babusci:2012ft}
D.~Babusci {\em et~al.} [KLOE Collaboration],
  \href{http://dx.doi.org/10.1016/j.physletb.2012.11.032}{Phys. Lett. B
  {\bfseries 718}, 910 (2013)}
  [\href{https://arxiv.org/abs/1209.4611}{{arXiv:1209.4611 [hep-ex]}}].

\bibitem{Arnaldi:2009aa}
R.~Arnaldi {\em et~al.} [NA60 Collaboration],
  \href{http://dx.doi.org/10.1016/j.physletb.2009.05.029}{Phys. Lett. B
  {\bfseries 677}, 260 (2009)}
  [\href{https://arxiv.org/abs/0902.2547}{{arXiv:0902.2547 [hep-ph]}}].

\bibitem{Aguar-Bartolome:2013vpw}
P.~Aguar-Bartolome {\em et~al.} [A2 Collaboration],
  \href{http://dx.doi.org/10.1103/PhysRevC.89.044608}{Phys. Rev. C {\bfseries
  89}, 044608 (2014)} [\href{https://arxiv.org/abs/1309.5648}{{arXiv:1309.5648
  [hep-ex]}}].

\bibitem{Arnaldi:2016pzu}
R.~Arnaldi {\em et~al.} [NA60 Collaboration],
  \href{http://dx.doi.org/10.1016/j.physletb.2016.04.013}{Phys. Lett. B
  {\bfseries 757}, 437 (2016)}
  [\href{https://arxiv.org/abs/1608.07898}{{arXiv:1608.07898 [hep-ex]}}].

\bibitem{Adlarson:2016hpp}
P.~Adlarson {\em et~al.} [A2 Collaboration],
  \href{http://dx.doi.org/10.1103/PhysRevC.95.035208}{Phys. Rev. C {\bfseries
  95}, 035208 (2017)}
  [\href{https://arxiv.org/abs/1609.04503}{{arXiv:1609.04503 [hep-ex]}}].

\bibitem{Hoferichter:2014vra}
M.~Hoferichter, B.~Kubis, S.~Leupold, F.~Niecknig, and S.~P. Schneider,
  \href{http://dx.doi.org/10.1140/epjc/s10052-014-3180-0}{Eur. Phys. J. C
  {\bfseries 74}, 3180 (2014)}
  [\href{https://arxiv.org/abs/1410.4691}{{arXiv:1410.4691 [hep-ph]}}].

\bibitem{Kubis:2015sga}
B.~Kubis and J.~Plenter,
  \href{http://dx.doi.org/10.1140/epjc/s10052-015-3495-5}{Eur. Phys. J. C
  {\bfseries 75}, 283 (2015)}
  [\href{https://arxiv.org/abs/1504.02588}{{arXiv:1504.02588 [hep-ph]}}].

\bibitem{Hanhart:2016pcd}
C.~Hanhart, S.~Holz, B.~Kubis, A.~Kup\'s\'c, A.~Wirzba, and C.-W. Xiao,
  \href{http://dx.doi.org/10.1140/epjc/s10052-017-4651-x}{Eur. Phys. J. C
  {\bfseries 77}, 98 (2017)}
  [\href{https://arxiv.org/abs/1611.09359}{{arXiv:1611.09359 [hep-ph]}}],
  [Erratum: Eur. Phys. J. C \textbf{78}, 450 (2018)].

\bibitem{Ablikim:2017fll}
M.~Ablikim {\em et~al.} [BESIII Collaboration],
  \href{http://dx.doi.org/10.1103/PhysRevLett.120.242003}{Phys. Rev. Lett.
  {\bfseries 120}, 242003 (2018)}
  [\href{https://arxiv.org/abs/1712.01525}{{arXiv:1712.01525 [hep-ex]}}].

\bibitem{Aubert:2007ef}
B.~Aubert {\em et~al.} [BaBar Collaboration],
  \href{http://dx.doi.org/10.1103/PhysRevD.76.092005}{Phys. Rev. D {\bfseries
  76}, 092005 (2007)} [\href{https://arxiv.org/abs/0708.2461}{{arXiv:0708.2461
  [hep-ex]}}], [Erratum: Phys. Rev. D \textbf{77}, 119902 (2008)].

\bibitem{TheBABAR:2018vvb}
J.~P. Lees {\em et~al.} [BaBar Collaboration],
  \href{http://dx.doi.org/10.1103/PhysRevD.97.052007}{Phys. Rev. D {\bfseries
  97}, 052007 (2018)}
  [\href{https://arxiv.org/abs/1801.02960}{{arXiv:1801.02960 [hep-ex]}}].

\bibitem{Achasov:2017kqm}
M.~N. Achasov {\em et~al.},
  \href{http://dx.doi.org/10.1103/PhysRevD.97.012008}{Phys. Rev. D {\bfseries
  97}, 012008 (2018)}
  [\href{https://arxiv.org/abs/1711.08862}{{arXiv:1711.08862 [hep-ex]}}].

\bibitem{Gribanov:2019qgw}
S.~S. Gribanov {\em et~al.} [CMD-3 Collaboration],
  \href{http://dx.doi.org/10.1007/JHEP01(2020)112}{JHEP {\bfseries 01}, 112
  (2020)} [\href{https://arxiv.org/abs/1907.08002}{{arXiv:1907.08002
  [hep-ex]}}].

\bibitem{Wess:1971yu}
J.~Wess and B.~Zumino,
  \href{http://dx.doi.org/10.1016/0370-2693(71)90582-X}{Phys. Lett. B
  {\bfseries 37}, 95 (1971)}.

\bibitem{Witten:1983tw}
E.~Witten, \href{http://dx.doi.org/10.1016/0550-3213(83)90063-9}{Nucl. Phys. B
  {\bfseries 223}, 422 (1983)}.

\bibitem{Omnes:1958hv}
R.~Omnès, \href{http://dx.doi.org/10.1007/BF02747746}{Nuovo Cim. {\bfseries
  8}, 316 (1958)}.

\bibitem{Fujikawa:2008ma}
M.~Fujikawa {\em et~al.} [Belle Collaboration],
  \href{http://dx.doi.org/10.1103/PhysRevD.78.072006}{Phys. Rev. D {\bfseries
  78}, 072006 (2008)} [\href{https://arxiv.org/abs/0805.3773}{{arXiv:0805.3773
  [hep-ex]}}].

\bibitem{Guo:2008nc}
F.-K. Guo, C.~Hanhart, F.~J. Llanes-Estrada, and U.-G. Mei{\ss}ner,
  \href{http://dx.doi.org/10.1016/j.physletb.2009.05.052}{Phys. Lett. B
  {\bfseries 678}, 90 (2009)}
  [\href{https://arxiv.org/abs/0812.3270}{{arXiv:0812.3270 [hep-ph]}}].

\bibitem{Hanhart:2012wi}
C.~Hanhart, \href{http://dx.doi.org/10.1016/j.physletb.2012.07.038}{Phys. Lett.
  B {\bfseries 715}, 170 (2012)}
  [\href{https://arxiv.org/abs/1203.6839}{{arXiv:1203.6839 [hep-ph]}}].

\bibitem{Zyla:2020zbs}
P.~A. Zyla {\em et~al.} [Particle Data Group],
  \href{http://dx.doi.org/10.1093/ptep/ptaa104}{PTEP {\bfseries 2020}, 083C01
  (2020)}.

\bibitem{Gasser:2018qtg}
J.~Gasser and A.~Rusetsky,
  \href{http://dx.doi.org/10.1140/epjc/s10052-018-6378-8}{Eur. Phys. J. C
  {\bfseries 78}, 906 (2018)}
  [\href{https://arxiv.org/abs/1809.06399}{{arXiv:1809.06399 [hep-ph]}}].

\bibitem{Niehus:2020gmf}
M.~Niehus, M.~Hoferichter, B.~Kubis, and J.~Ruiz~de Elvira,
  \href{http://dx.doi.org/10.1103/PhysRevLett.126.102002}{Phys. Rev. Lett.
  {\bfseries 126}, 102002 (2021)}
  [\href{https://arxiv.org/abs/2009.04479}{{arXiv:2009.04479 [hep-ph]}}].

\bibitem{Akhmetshin:2000wv}
R.~R. Akhmetshin {\em et~al.} [CMD-2 Collaboration],
  \href{http://dx.doi.org/10.1016/S0370-2693(00)00937-0}{Phys. Lett. B
  {\bfseries 489}, 125 (2000)}
  [\href{https://arxiv.org/abs/hep-ex/0009013}{{arXiv:hep-ex/0009013}}].

\bibitem{Dai:2013joa}
L.-Y. Dai, J.~Portoles, and O.~Shekhovtsova,
  \href{http://dx.doi.org/10.1103/PhysRevD.88.056001}{Phys. Rev. D {\bfseries
  88}, 056001 (2013)} [\href{https://arxiv.org/abs/1305.5751}{{arXiv:1305.5751
  [hep-ph]}}].

\bibitem{Volkov:2013zba}
M.~K. Volkov, A.~B. Arbuzov, and D.~G. Kostunin,
  \href{http://dx.doi.org/10.1103/PhysRevC.89.015202}{Phys. Rev. C {\bfseries
  89}, 015202 (2014)} [\href{https://arxiv.org/abs/1310.8484}{{arXiv:1310.8484
  [hep-ph]}}].

\bibitem{Qin:2020udp}
W.~Qin, L.-Y. Dai, and J.~Portoles,
  \href{http://dx.doi.org/10.1007/JHEP03(2021)092}{JHEP {\bfseries 03}, 092
  (2021)} [\href{https://arxiv.org/abs/2011.09618}{{arXiv:2011.09618
  [hep-ph]}}].

\bibitem{Achasov:1984ru}
N.~N. Achasov and V.~A. Karnakov, JETP Lett. {\bfseries 39}, 342 (1984).

\bibitem{Lomon:2012pn}
E.~L. Lomon and S.~Pacetti,
  \href{http://dx.doi.org/10.1103/PhysRevD.86.039901}{Phys. Rev. D {\bfseries
  85}, 113004 (2012)} [\href{https://arxiv.org/abs/1201.6126}{{arXiv:1201.6126
  [hep-ph]}}], [Erratum: Phys. Rev. D \textbf{86}, 039901 (2012)].

\bibitem{Moussallam:2013una}
B.~Moussallam, \href{http://dx.doi.org/10.1140/epjc/s10052-013-2539-y}{Eur.
  Phys. J. C {\bfseries 73}, 2539 (2013)}
  [\href{https://arxiv.org/abs/1305.3143}{{arXiv:1305.3143 [hep-ph]}}].

\bibitem{VonHippel:1972fg}
F.~von Hippel and C.~Quigg,
  \href{http://dx.doi.org/10.1103/PhysRevD.5.624}{Phys. Rev. D {\bfseries 5},
  624 (1972)}.

\bibitem{Beladidze:1993km}
G.~M. Beladidze {\em et~al.} [VES Collaboration],
  \href{http://dx.doi.org/10.1016/0370-2693(93)91224-B}{Phys. Lett. B
  {\bfseries 313}, 276 (1993)}.

\bibitem{Farrar:1975yb}
G.~R. Farrar and D.~R. Jackson,
  \href{http://dx.doi.org/10.1103/PhysRevLett.35.1416}{Phys. Rev. Lett.
  {\bfseries 35}, 1416 (1975)}.

\bibitem{Vainshtein:1977db}
A.~I. Vainshtein and V.~I. Zakharov,
  \href{http://dx.doi.org/10.1016/0370-2693(78)90140-5}{Phys. Lett. B
  {\bfseries 72}, 368 (1978)}.

\bibitem{Lepage:1979zb}
G.~P. Lepage and S.~J. Brodsky,
  \href{http://dx.doi.org/10.1016/0370-2693(79)90554-9}{Phys. Lett. B
  {\bfseries 87}, 359 (1979)}.

\bibitem{Lepage:1980fj}
G.~P. Lepage and S.~J. Brodsky,
  \href{http://dx.doi.org/10.1103/PhysRevD.22.2157}{Phys. Rev. D {\bfseries
  22}, 2157 (1980)}.

\bibitem{Guo:2011ir}
F.-K. Guo, B.~Kubis, and A.~Wirzba,
  \href{http://dx.doi.org/10.1103/PhysRevD.85.014014}{Phys. Rev. D {\bfseries
  85}, 014014 (2012)} [\href{https://arxiv.org/abs/1111.5949}{{arXiv:1111.5949
  [hep-ph]}}].

\bibitem{Ablikim:2014eoc}
M.~Ablikim {\em et~al.} [BESIII Collaboration],
  \href{http://dx.doi.org/10.1103/PhysRevLett.112.251801}{Phys. Rev. Lett.
  {\bfseries 112}, 251801 (2014)}
  [\href{https://arxiv.org/abs/1404.0096}{{arXiv:1404.0096 [hep-ex]}}],
  [Addendum: Phys.Rev.Lett. 113, 039903 (2014)].

\bibitem{Caprini:2011ky}
I.~Caprini, G.~Colangelo, and H.~Leutwyler,
  \href{http://dx.doi.org/10.1140/epjc/s10052-012-1860-1}{Eur. Phys. J. C
  {\bfseries 72}, 1860 (2012)}
  [\href{https://arxiv.org/abs/1111.7160}{{arXiv:1111.7160 [hep-ph]}}].

\bibitem{Gasser:1983yg}
J.~Gasser and H.~Leutwyler,
  \href{http://dx.doi.org/10.1016/0003-4916(84)90242-2}{Annals Phys. {\bfseries
  158}, 142 (1984)}.

\bibitem{Dax:2018rvs}
M.~Dax, T.~Isken, and B.~Kubis,
  \href{http://dx.doi.org/10.1140/epjc/s10052-018-6346-3}{Eur. Phys. J. C
  {\bfseries 78}, 859 (2018)}
  [\href{https://arxiv.org/abs/1808.08957}{{arXiv:1808.08957 [hep-ph]}}].

\bibitem{Aoki:2019cca}
S.~Aoki {\em et~al.} [FLAG Collaboration],
  \href{http://dx.doi.org/10.1140/epjc/s10052-019-7354-7}{Eur. Phys. J. C
  {\bfseries 80}, 113 (2020)}
  [\href{https://arxiv.org/abs/1902.08191}{{arXiv:1902.08191 [hep-lat]}}].

\bibitem{Bazavov:2010hj}
A.~Bazavov {\em et~al.} [MILC Collaboration],
  \href{http://dx.doi.org/10.22323/1.105.0074}{PoS {\bfseries LATTICE2010}, 074
  (2010)} [\href{https://arxiv.org/abs/1012.0868}{{arXiv:1012.0868
  [hep-lat]}}].

\bibitem{Borsanyi:2012zv}
S.~Bors\'anyi, S.~D{\"u}rr, Z.~Fodor, S.~Krieg, A.~Sch{\"a}fer, E.~E. Scholz,
  and K.~K. Szab\'o, \href{http://dx.doi.org/10.1103/PhysRevD.88.014513}{Phys.
  Rev. D {\bfseries 88}, 014513 (2013)}
  [\href{https://arxiv.org/abs/1205.0788}{{arXiv:1205.0788 [hep-lat]}}].

\bibitem{Durr:2013goa}
S.~D\"urr {\em et~al.} [BMW Collaboration],
  \href{http://dx.doi.org/10.1103/PhysRevD.90.114504}{Phys. Rev. D {\bfseries
  90}, 114504 (2014)} [\href{https://arxiv.org/abs/1310.3626}{{arXiv:1310.3626
  [hep-lat]}}].

\bibitem{Boyle:2015exm}
P.~A. Boyle {\em et~al.},
  \href{http://dx.doi.org/10.1103/PhysRevD.93.054502}{Phys. Rev. D {\bfseries
  93}, 054502 (2016)}
  [\href{https://arxiv.org/abs/1511.01950}{{arXiv:1511.01950 [hep-lat]}}].

\bibitem{Beane:2011zm}
S.~R. Beane, W.~Detmold, P.~M. Junnarkar, T.~C. Luu, K.~Orginos, A.~Parre\~no,
  M.~J. Savage, A.~Torok, and A.~Walker-Loud,
  \href{http://dx.doi.org/10.1103/PhysRevD.86.094509}{Phys. Rev. D {\bfseries
  86}, 094509 (2012)} [\href{https://arxiv.org/abs/1108.1380}{{arXiv:1108.1380
  [hep-lat]}}].

\bibitem{Dobado:1989qm}
A.~Dobado, M.~J. Herrero, and T.~N. Truong,
  \href{http://dx.doi.org/10.1016/0370-2693(90)90109-J}{Phys. Lett. B
  {\bfseries 235}, 134 (1990)}.

\bibitem{Truong:1991gv}
T.~N. Truong, \href{http://dx.doi.org/10.1103/PhysRevLett.67.2260}{Phys. Rev.
  Lett. {\bfseries 67}, 2260 (1991)}.

\bibitem{Dobado:1992ha}
A.~Dobado and J.~R. Pel\'aez,
  \href{http://dx.doi.org/10.1103/PhysRevD.47.4883}{Phys. Rev. D {\bfseries
  47}, 4883 (1993)}
  [\href{https://arxiv.org/abs/hep-ph/9301276}{{arXiv:hep-ph/9301276}}].

\end{thebibliography}\endgroup

\end{document}